\documentclass[acmsmall, nonacm]{acmart}

\AtBeginDocument{%
  \providecommand\BibTeX{{%
    \normalfont B\kern-0.5em{\scshape i\kern-0.25em b}\kern-0.8em\TeX}}}

\usepackage[table]{xcolor}
\usepackage{colortbl}
\usepackage{multirow}
\usepackage{makecell}
\usepackage{tabularx}
\usepackage{longtable}
\usepackage{booktabs}
\usepackage{array}
\usepackage{float}

\setcopyright{rightsretained}
\usepackage[parfill]{parskip}
\usepackage[table]{xcolor}

\begin{document}

\title{New Election, Same Old Story: How Tropes Form a Grammar of Election Rumors}

\author{Stephen Prochaska}
\authornote{Both authors contributed equally to this research.}
\affiliation{%
  \institution{Center for an Informed Public, Information School, University of Washington}
  \city{Seattle}
  \country{USA}}
\email{sprochas@uw.edu}

\author{Mert Can Bayar}
\authornotemark[1]
\affiliation{
  \institution{Center for an Informed Public, Interdisciplinary Arts and Sciences, University of Washington Bothell}
  \city{Seattle}
  \country{USA}
}

\author{Ashlyn B. Aske}
\affiliation{%
  \institution{Center for an Informed Public, University of Washington School of Law}
  \city{Seattle}
  \country{USA}
}

\author{Joseph Schafer}
\affiliation{%
  \institution{Center for an Informed Public, Human Centered Design \& Engineering, University of Washington}
  \city{Seattle}
  \country{USA}
}

\author{Kate Starbird}
\email{kstarbi@uw.edu}
\affiliation{%
  \institution{Center for an Informed Public, Human Centered Design \& Engineering, University of Washington}
  \city{Seattle}
  \state{Washington}
  \country{USA}
}

\renewcommand{\shortauthors}{Stephen Prochaska et al.}  

\begin{abstract}
Recent elections in the United States have seen online rumors become a regular feature of election discourse as voters and election officials work to navigate a sea of digitally spread claims and propaganda. For the uninitiated, these rumors can feel overwhelming and random, sowing doubt about the integrity and legitimacy of recent elections. In this paper, we reviewed and qualitatively coded a large dataset of 590 election rumors collected during the the 2020 and 2022 national elections. We found that there are tropes that recur time and time again in election rumors, highlighting how these rumors are not as random as they may first appear. We make two primary contributions in this work: first, we introduce a hierarchical framework that structures observed tropes in three tiers (Meta-category, Meso-trope, and Micro-trope). Second, combining the framework with our coded data, we identify a visible grammar of election rumors, highlighting how in the majority of cases a rumor contains an Actor, and Object, and an Action, with optional grammatical components of Suggested Remedies and Scene \& Setting. The framework and grammar can be used by other researchers to act as the underlying information architecture of analyses examining similar rumors, enabling the rapid identification, analysis, and prediction of election rumors.
\end{abstract}


\begin{CCSXML}
<ccs2012>
   <concept>
       <concept_id>10003120.10003130.10011762</concept_id>
       <concept_desc>Human-centered computing~Empirical studies in collaborative and social computing</concept_desc>
       <concept_significance>300</concept_significance>
       </concept>
 </ccs2012>
\end{CCSXML}

\ccsdesc[300]{Human-centered computing~Empirical studies in collaborative and social computing}

\keywords{Trope, Rumor, Social Media, Elections, United States,  Misinformation, Disinformation, Storytelling, Narrative, Story}

\maketitle

\section{Introduction}
On November 4th, 2020, an angry mob gathered outside the TCF Center, a vote counting facility in Detroit, Michigan. As poll workers inside attempted to tabulate ballots, protesters pounded on windows, chanting “stop the count!” \cite{reillyHowTrumpAllies2024} Tensions escalated so much that security covered the facility’s windows to protect poll workers and prevent spectators from viewing private voter information \cite{darcyRightwingMediaPortrayed}. Within hours, the Trump campaign filed a lawsuit requesting a judge “stop the count” in Michigan. The chaos and confusion gained national attention and fed into the broader “rigged election” narrative that came to define the aftermath of the 2020 election.

The events at the TCF Center were driven by rumors shared and spread on social media \cite{reillyHowTrumpAllies2024}. They began with an eyewitness video, taken in the early morning hours after the election, showing a man unloading a black case from a van. The videographer remarked that the case looked suspiciously like a lockbox used to transport ballots. Although later reporting revealed that the case contained a newscaster’s camera equipment, rumors of fake ballots being brought into the TCF Center under the cover of darkness quickly took off. Later that day, growing crowds of observers gathered at the facility, and the covering of its windows further fueled allegations of a cover up. To audiences already primed to interpret irregularities as evidence of election fraud, the black case, the early-morning timing, the crowds of observers, and the covered windows came together to support a story of a stolen election.

Rumors about election administration and vote counting — and mobilization around those rumors through protests and lawsuits — were not unique to the TCF Center or to Michigan. Similar rumors emerged around the country. Over time, and with the help of political influencers and organizers \cite{starbirdInfluenceImprovisationParticipatory2023d}, these rumors coalesced into a broad and persistent conspiracy theory about a stolen election and contributed to efforts to contest the results of the 2020 election, including during the events of January 6, 2021. This conspiracy theory continues to be repeated by President Trump and his supporters as we head into the 2026 U.S. midterms \cite{ericksonTrumpClaims20202026} and used to motivate and justify changes to election laws \cite{singhTrumpAdministrationsCampaign2025, mcdonaldFactCheckingTrumpsElection2026}. Similar rumors and “rigged election” narratives have emerged in subsequent U.S. elections, most recently alongside false claims of election rigging in the June 2026 California primary. Rumoring about election integrity isn’t confined to the U.S., and similar rumors are spreading around elections in other democratic countries, including Brazil \cite{bastosInsurrectionistPlaybookJair2023} and South Korea \cite{leePoliticalDiscourseTwitter2024}. These rumors have real world consequences, leading to contested election results, diminished trust in democracy, and cynical changes to election laws that make voting processes less free and fair. Rumors about voting and election administration — specifically "delegitimizing rumors” that function to sow distrust in election processes and results — are a growing threat to democracy in the U.S. and in other countries around the world.

Our research team has conducted extensive research into rumors about U.S. elections since 2020. Working across three election cycles and heading into a fourth, it has become clear that rumors about voting and election administration are neither original nor are they random. Allegations of “slow vote counting”, “suitcases of ballots”, and “cover ups” may have seemed relatively novel in 2020, but they grew familiar when we saw them again in 2022, in 2024, and in early primaries in the 2026 cycle. We observed there are common elements or pieces of rumors that repeat over and over again, even as the specific events around them change.

To understand what these patterns are, and what they reveal about how election rumors take shape and why they spread, our research turns to the concept of a narrative trope. Tropes are the building blocks of stories, recurring narrative elements that allow audiences to situate new events within existing cultural and political narratives by surfacing recognizable actors, actions, objects, and scenes. For example, in the case of rumors surrounding the Detroit TCF Center in 2020, eyewitness reports of “suitcases of ballots” (object trope) delivered to “a vote-counting site” (scene/setting trope) in “the middle of the night” (scene/setting trope) followed by a “cover-up” (action trope) by “election officials” (character trope) contribute to a larger story of election fraud. Mapping out the common election rumor tropes and illuminating how they co-occur and relate to real-world events can help us understand how and why certain rumors capture online attention and drive offline mobilization.

In this research, we set out to identify and map the salient tropes of rumors about U.S. election administration in 2020 and 2022. We highlight how tropes are not used randomly, instead identifying a grammar for when and how particular types of tropes come together to form election rumors. We build upon prior work \cite{kennedyRepeatSpreadersElection2022d, schaferElectionRumors2022DatasetElection2025} that identified 590 distinct rumors about election integrity spreading around the 2020 and 2022 elections and a curated dataset of 34.2M tweets with search terms associated with those rumors. We first identify all of the major tropes that appeared in those rumors, blending computational and qualitative analysis to develop a multi-level coding scheme that organizes these tropes according to their narrative function. We then map all of the tropes onto the different rumors that contained them, and analyze those mappings for frequency and patterns of co-occurence.

This work lays out an enumeration of common tropes within rumors about U.S. election administration and presents a three-tiered framework for grouping specific tropes into higher-level categories. Within this framework, we introduce a proposed “grammar” of election rumors, highlighting how the vast majority of rumors follow a loose set of rules in their composition. Most rumors in our data revolved around a core composed of actors, objects, and actions taken by or involving those actors and objects. Like linguistic grammatical rules, this grammar is sometimes broken by cases of “improper grammar,” but it nevertheless provides a useful starting point for understanding the common structure of election rumors. In the election context, we believe that these identified tropes, their structure, and the broader rumor trope framework can help researchers and communicators more quickly identify emerging rumors and anticipate how they may develop. These insights could also inform predictive models aimed at anticipating how new rumors emerge from novel events. Beyond the election context, we hope this work will help others better understand the underlying structure—or grammar—of rumors in other domains and the relationship between novel events and familiar tropes.

\section{Background}
\subsection{Rumoring and Election Rumoring}

Rumors are unofficial stories that spread through informal channels \cite{allportPsychologyRumor1947, kapfererRumorsUsesInterpretations2013, shibutaniImprovisedNewsSociological1966}. Prior research shows that they often emerge in response to uncertain or ambiguous events, particularly when demand for information is high but trusted, reliable information is limited \cite{shibutaniImprovisedNewsSociological1966}. Shibutani emphasizes the interactive nature of this process: rumors are not independently existing objects but are co-constructed by communities through iterative discourse. Rumors also become more prevalent and spread more widely when uncertainty is paired with higher stakes for the communities in which they circulate \cite{allportPsychologyRumor1947, spiroRumorsHaveRules2023a}, a relationship also observed in recent work on election misinformation \cite{wackLegislatingUncertaintyElection2026}. Research on online rumoring has therefore often focused on periods of crisis and uncertainty, including the 2013 Boston Marathon bombing \cite{starbirdRumorsFalseFlags2014a, arifHowInformationSnowballs2016} and the 2010 Deepwater Horizon oil spill \cite{spiroRumoringExtremeEvents2012a}. But rumors also spread in less clearly bounded situations of prolonged uncertainty, including reactions to economic and regulatory changes in an informal marketplace in India \cite{chandraRumorsCollectiveSensemaking2019b}.
Increasingly, rumors surrounding elections, events that are both uncertain and high-stakes, spread in both online and offline spaces \cite{davidsonAUDIENCEENGAGEMENTDYNAMICS2024, prochaskaMobilizingManufacturedReality2023b}. In particular, since the 2020 election in the United States, online rumors have become a recurring feature of U.S. elections, requiring election administrators to dedicate time and labor to responding and communicating about election processes that are frequent subjects of rumoring \cite{grossWhatHappensWhen2023}. Online election rumors have mixed with disinformation campaigns and propaganda \cite{benklerMailInVoterFraud2020b, prochaskaMobilizingManufacturedReality2023b} and online attention dynamics \cite{starbirdWhatGoingEvidenceframe2025b, starbirdInfluenceImprovisationParticipatory2023d} to produce hundreds of distinct rumors about voting, election administration, and election integrity.
Research examining online election rumors and associated disinformation has documented how rumors spreading within a community often rely on a shared, prior body of knowledge, meanings, and epistemic approach within a community that is leveraged by influencers and audiences alike \cite{marwickDisinformationCulturalNarrative2026a,maThinkingThemselvesExamining2023}. In particular, narratives often provide structure for otherwise disparate or ambiguous claims, tying individual rumors together as part of a larger whole \cite{marwickDisinformationCulturalNarrative2026a}. Prior work has also highlighted how audiences and influencers engage in collaborative, improvisational storytelling that facilitates the spread of false and misleading election rumors, often capitalizing on the online attention economy to do so \cite{starbirdInfluenceImprovisationParticipatory2023d, prochaskaMobilizingManufacturedReality2023b}. 
Prior research emphasizes that narrative is one of the central features of both rumors and disinformation. Here, we offer a more granular view of how stories take shape and become part of broader narratives through recurring objects, actors, actions, and themes—that is, through tropes. Tropes are building blocks of storytelling because they make new stories feel familiar and recognizable \cite{busseFanFictionTropes2017, thompsonMotifIndexFolkLiterature1946}, a function that appears to operate similarly in the election rumors we examine \cite{spiroRumorsHaveRules2023a}. Our framework focuses primarily on these component-level tropes rather than broader plot tropes. Election rumors in our data already share a relatively common overarching plot centered on election processes going wrong, being manipulated, or being interpreted as fraudulent; what varies across rumors are the actors, objects, actions, settings, and remedies through which that broader story is told.

\subsection{Narrative Tropes}

Tropes have been studied in a variety of disciplines, but we draw most directly from media and film studies, where tropes are understood as recurring patterns—such as narrative arcs, themes, character types, settings, or plot points—that make works within a genre more legible. Many scholars use related terms such as motifs, archetypes, and formulas for overlapping concepts. These tropes afford storytellers (such as authors and screenwriters) a form of shorthand, where repeated figures, locations, actions, and meanings can communicate association with broader, recurring themes.\footnote{Note, research in other humanistic fields, including rhetoric, philosophy, and linguistics, also studies tropes as particular linguistic devices, such as metaphor, metonymy, synecdoche, and irony \cite{burkeGrammarMotives1969}; however, this lineage of research is separate and not our focus in this study.}

Many scholars have discussed tropes without necessarily using the term itself. Vladimir Propp \cite{proppMorphologyFolktale1968} is one of the earliest examples of this type of structural analysis of narrative subunits. In his work on Russian fairy tales, he argues that many stories share a common structure based on recurring “functions” or narrative units, which are the stable, essential building blocks of the plot regardless of the characters or specific details. Propp goes further to identify recurring roles such as villains and heroes. Similarly, Thompson defines the “motif” as the “smallest element in a tale having a power to persist in tradition,” emphasizing how recurring story elements endure by carrying cultural meanings and collective memories across time. Umberto Eco \cite{ecoCasablancaCultMovies1985} makes a related point when he notes that while one cliché makes us laugh, a hundred clichés move us. In his analysis of ‘Casablanca’, he argues that the film’s power comes in part from the dense layering of familiar narrative figures and situations. These archetypes, motifs, and tropes are what we call the building blocks of stories: stories change, but they also stay the same, and it is the persistence of these recurring elements that makes them recognizable, relatable, persuasive, and often compelling to follow.

Tropes in election rumors are similar in this sense to fan fiction, where communities engage in storytelling through repeated and recognizable narrative patterns. Rumoring audiences, influencers, and sensemakers build stories out of elite-led genres and meta-narratives of untrustworthy elections, election fraud, and conspiracy \cite{benklerMailInVoterFraud2020b, starbirdInfluenceImprovisationParticipatory2023d}. This follows a familiar path of narrative creation through the repetition of themes and tropes. Many election rumors are organized around a narrative structure similar to what Christopher Booker calls “overcoming the monster,” where heroes face a villain or series of villains—in our case, the deep state, election officials, tech companies, Democrats, or the Trump administration—and the rumors attempt to expose this plot, awaken the masses, and mobilize them to fight back. In turn, “overcoming the monster” requires some sort of counteraction: a strategy to deal with the threat or restore order. Michel de Certeau, in his study of everyday life, characterizes stories as “repertoires of schemas of action” \cite{decerteauPracticeEverydayLife1988, tangherliniGenerativeModelLegend2018}, implicitly recognizing that stories model how one might react to various events and the range of potential outcomes of those reactions. Tangherlini \cite{tangherliniGenerativeModelLegend2018, tangherliniParlerGamesNarrative2025} similarly proposes an actant-interaction model in which stories are built through recurring relations among people, places, and things, and argues that stable narrative frameworks emerge as storytellers repeatedly select from a familiar pool of actants and relationships.\footnote{Similar to our framework, Tangherlini also proposes three layers of story structure at macro, meso, and micro levels. His focus, however, is more on the functions and narrative spaces through which legend stories are organized, whereas our three-layer framework is structured through the skeleton of a rumor—its actors, actions, and objects—rather than the broader deep stories or cultural settings in which those stories circulate.} Conspiracy theories often work similarly by identifying a hidden threat or secret plot, but they do not always offer a concrete plan of action \cite{bayarConspiracistTheoryPower2025}. Election conspiracy theories and rumors, however, often fill this gap by not only identifying the threat but also pointing to ways believers can mobilize against it \cite{bayarPartisanNonpartisanConspiracy2025}. This is where Tangherlini’s and our framework’s emphasis on ‘call to action’ (or suggested remedies) becomes especially important.

Recent research has increasingly focused on repeated tropes, narratives, and patterns on social media within propaganda, misinformation, and disinformation \cite{blackburnCorpusDevelopmentStudying2020, hughesDevelopmentCodebookOnline2021b, liClassifyingCOVID19Vaccine2023, nivenTropeBasedComputationalApproach2026a}. Some of this work has moved beyond mere facticity, arguing that misleading content often works not simply by making false claims, but by embedding those claims within recurring narratives that are recognizable, persuasive, and identity-laden \cite{reddiIdentityPropagandaRacial2023b, marwickDisinformationCulturalNarrative2026a}. Related work has used computational methods to identify narrative frameworks, actants, and relationships at scale \cite{tangherliniAutomatedPipelineDiscovery2020a, tangherliniParlerGamesNarrative2025}. 

Our work builds on this literature by adapting studies of narrative, propaganda, and recurring tropes to rumor-based election discourse on Twitter/X through an actor-, object-, and action-centered framework, using qualitative coding of rumors to identify recurring patterns and develop a theory-driven framework of election-rumor tropes.

\section{Methods}
In this study, we used a multi-step, mixed method approach to map out common narrative tropes related to the 2020 and 2022 U.S. national elections. We combined qualitative coding with computational tools to make analysis of a much larger dataset feasible. Specifically, a team of coders with extensive expertise in the online spread of U.S. election rumors iteratively developed and applied a coding scheme to 590 (568 after removing unrelated or difficult to understand rumors) rumors identified in prior work \cite{kennedyRepeatSpreadersElection2022d, schaferElectionRumors2022DatasetElection2025}.

\subsection{Data}

The data for this study draws on rumors extracted from two broader corpora of election-related Twitter discourse. The first corpus, covering the 2020 Presidential election cycle, is documented in Kennedy et al. \cite{kennedyRepeatSpreadersElection2022d}; the second, focused on 2022 Midterm elections, is described in Schafer \& Duskin et al. \cite{schaferElectionRumors2022DatasetElection2025}. Each corpus was built using Twitter’s streaming API, capturing tweets around keywords tied to electoral processes, such as “ballot” and “election,” as well as language associated with false claims around election “fraud.” Data collection for the 2020 corpus spanned August through December 2020, ultimately producing more than a billion tweets about the election, of which 32.4 million were organized into 455 discrete rumors \cite{kennedyRepeatSpreadersElection2022d}. The 2022 corpus was gathered from roughly July 2022 to March 2023 and similarly exceeded a billion tweets in total volume. From this collection, a refined subset of approximately 1.8 million tweets was mapped to 135 individual rumors occurring between September and December 2022. We initially identified 590 rumors across the two datasets. Of these, 23 were excluded because they were outside the scope of the study or did not contain enough associated tweets to support meaningful rumor-level summaries. This left 568 rumors in the final dataset. 

Following Prochaska et al. \cite{prochaskaDeepStorytellingCollective2025b}, we coded our data at the level of rumors — collections of 10s to 100,000s of tweets associated with a specific claim or story — rather than individual tweets. Researchers relied upon qualitative analysis of three rumor summarization tools (LLM-assisted summaries, word clouds, and samples of both random and the most-retweeted tweets) to identify the tropes within each rumor. The most central tool in this process was the LLM-assisted rumor summary.

We generated AI-assisted summaries of each rumor using ChatGPT 4.1 mini.\footnote{We created the summaries in May 2025 accessing ChatGPT 4.1 mini through OpenAI’s API.} These summaries were based on up to 1,000 tweets:\footnote{The actual number of tweets per-rumor varied due to the variation in number of tweets making up a given rumor.} the 500 most retweeted tweets and 500 randomly selected tweets.\footnote{The first 500 was intended to simulate the type of virality most common on Twitter: large concentrations of retweets for a fairly small set of influential tweets. To capture this dynamic, we included up to 10 copies each of the top 50 most retweeted tweets in each rumor. To capture themes that may have been absent from this subset, the second half also included a random sample of 500 tweets (excluding retweets).} They were produced through a two-step hierarchical process, similar to recent LLM-based summarization workflows \cite{ouContextAwareHierarchicalMerging2025, chhikaraLaMSUMAmplifyingVoices2024}. First, the LLM summarized batches of tweets, identifying the specific themes and claims present in each batch. Second, it synthesized the batch-level summaries associated with each rumor into a single comprehensive rumor-level summary. To help capture narrative tropes, we prompted the LLM to focus on key elements of each rumor, including the actors or entities named, the specific claims being made, and the actions being alleged. This process produced 568 rumor summaries (a summary per rumor), which served as the basis for identifying recurring tropes across the dataset.

\subsection{Developing the Coding Scheme}

Our research team used an iterative, inductive approach to identify the tropes and develop the coding scheme. As we reviewed initial rumor descriptions, tweets, and summaries, members of the research team noted recurring themes, attending in particular to objects, such as “mail-in ballots,” and other entities, such as “election officials” as villain actors, that repeatedly appeared across multiple rumors. Through this process of exploration and open coding, we identified a three-part structure in the rumor data: meta categories, meso tropes, and micro tropes. The meta categories identify the broad narrative components within rumors, including Actors, Actor-Action, Object, Object-Action, Scene and Setting, and Suggested Remedies. The meso tropes identify broader types of actors, objects, actions, scenes, and remedies, such as villain as a type of actor or foreign interference as a type of actor-action. The micro tropes identify more specific narrative patterns within these broader types, such as election officials as a micro trope under the villain meso category, or influence operation as a micro trope under foreign interference. The full coding scheme is included in Appendix A, and a subset of categories is shown in Figure~\ref{fig:wheel_chart}.

We organize part of the framework around Actor-Action and Object-Action categories rather than treating “action” as a separate meta-category. This choice reflects the storytelling structure of election rumors, where actions often gain meaning only through their attachment to particular actors or objects. Rumors are narrative forms \cite{kapfererRumorsUsesInterpretations2013}, and the tropes that recur within them are often not isolated actions but actor-action or object-action combinations. For example, “ballot harvesting” is not simply an action detached from an object; separating it into “ballot” and “harvesting” would produce broad, ambiguous components while obscuring the recognizable trope itself. This is especially important in social media discourse, where rumoring is shaped by performance, in-group cues, and online attention economies \cite{tomsonInfluenceWhatsReal, starbirdInfluenceImprovisationParticipatory2023d, prochaskaDeepStorytellingCollective2025b}. Participants often assume audiences already know which actors or objects are implicated, so the action can become the focal point while the actor or object remains implied. By coding these patterns as Actor-Action and Object-Action, the framework preserves the recognizable narrative unit of the rumor while avoiding overly fragmented codes that would be less useful for analysis. This distinction is one of the more distinctive features of our framework, separating it from other similar typologies. It allows us to capture how rumors are communicated on shared knowledge, often leaving important details (actors or objects) unnamed or implied. 

In this framework, we classified recurring themes as meso or micro tropes depending on their specificity and narrative function. For instance, poll workers often appeared as villains in the rumor data. We therefore coded poll workers as a micro trope under the villain meso trope, which is itself located within the broader Actors meta category. By contrast, rumors about hacked voting machines were coded as a micro trope under the manipulated machines meso trope. This meso category also included corrupt voting machines, with “hacked” referring to claims that machines were manipulated after their design, and “corrupt” referring to claims that the machines were designed for fraud from the beginning. In this case, manipulated machines functioned as the nested meso category under which the hacked and corrupt machine tropes belonged.

Similarly, we coded foreign interference as a meso trope because rumors about foreign interference included several more specific subcategories. These included claims about influence operations, machine manipulation, manipulated ballots, or more general claims of corruption based on someone’s or some company’s alleged association with a foreign adversary. For example, some versions of the Dominion conspiracy theory tied voting machines to China based on trade ties. In cases like this, foreign interference operated as a broader meso trope that organized several more specific micro tropes.

As we added new tropes to the coding scheme and iterated on existing ones, we simultaneously began testing the initial codes on rumor summaries. Each coder was assigned the same randomly selected set of 50 rumors. We then refined the codes through repeated discussion, calibrating a shared understanding of when specific codes should be applied. To guide these conversations, we focused on achieving coder consistency and focused especially on codes with low coder agreement, working toward a relatively consistent and replicable consensus in how the codes were applied. As we approached high coder consistency, we began applying the framework to the full dataset with a nearly complete version of the coding scheme.\footnote{During calibration, we updated existing tropes if deemed necessary and added several new ones as they emerged from the coding process. Each coder then revisited their assigned rumors and updated their codes to reflect the revised coding scheme.}

\begin{figure}[htbp]
    \centering
    \textbf{Methods Overview Schematic}\\[0.5em]
    \includegraphics[width=0.8\textwidth]{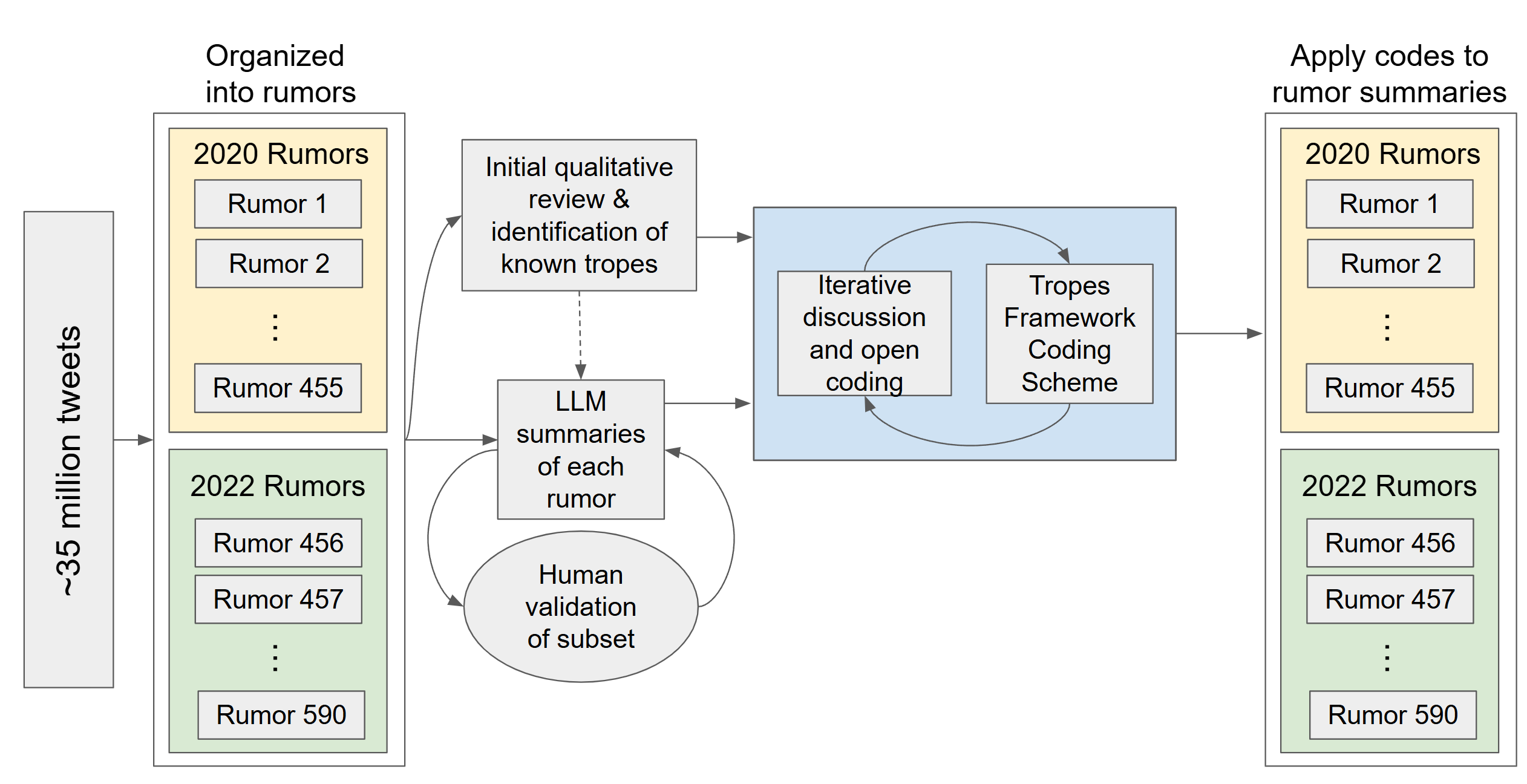}
    \caption{Visualization of methodological workflow. We began by exploring the rumors and their underlying tweets through an initial qualitative review of the data, which helped us identify recurring trope patterns. We then moved into a more systematic coding process using AI-generated rumor summaries, iterating on the scheme until we achieved a high level of coder convergence. Through this iterative calibration, we aligned the coding scheme with the underlying dimensional structure of the framework, meta, meso, and micro categories, and refined it until we reached a high level of consistency. This process allowed us to finalize the trope framework and establish the coding scheme.}
    \Description{Flowchart showing the evolution of research phases. It starts on the left with a grey box labelled "~35 million tweets", which connects to a box to the right labelled "Organized into rumors." Within this box are smaller boxes labelled "Rumor 1", "Rumor 2", etc. showing rumors Rumors 1 through 455 came from 2020 while rumors 456 through 590 came from 2022. This data box connects to the right to two parallel boxes, one indicating qualitative review of the data and the other LLM summarization of that same data. The qualitative review informed the LLM summarization, which is shown in the visual by a connecting arrow. These then connect to a set of boxes showing how the coding scheme emerged from iterative discussion and coding, and was finally applied to the rumor summaries of the original dataset.}
    \label{fig:methods_visual}
\end{figure}

\subsection{Applying the Coding Scheme}

During calibration on a shared set of 50 rumors, coders discussed disagreements and refined code definitions before final coding. We used remaining disagreements to further refine the coding scheme and develop shared definitions before moving to final coding. Then, we began applying the framework to systematically code tropes within all of the U.S. 2020/2022 Election rumors. Our team of three researchers coded each rumor, identifying each of the micro and meso tropes by inspecting the rumor summaries alongside word clouds drawn from the tweets that constituted each rumor. Coders could select multiple micro tropes for the same rumor within a given meso category. Each rumor was randomly assigned to two coders. During the coding process, the team of three coders met weekly to discuss challenges, ask questions, resolve ambiguities, and revise the coding scheme when necessary.\footnote{ We tried to keep the coding scheme unchanged after we had started coding the whole dataset. We revised the coding scheme only if there was an important trope that we missed or a conceptual change we had to make on an existing trope such as broadening or restricting a trope’s scope.}

After applying codes to the entire dataset, we conducted a two-step process for resolving disagreements. First, we arbitrated disagreements using a third coder (who was not one of the original coders), deciding which code, or set of codes, best captured the rumor in question. During this process, each arbitrator marked codes or rumors that were difficult to resolve and brought them back to the group for consensus coding. We eliminated 23 rumors that all coders agreed were out of scope of the project, either because the rumor was only vaguely connected to election administration, or because the rumor summary was too ambiguous to reliably interpret. This resulted in a final coded dataset consisting of 568 rumors which we used to calculate the code frequencies reported below.

\section{Findings}
In this section, we present the final products resulting from the meticulous process of exploring, coding, and framing the dataset. First and foremost, the primary contribution is the tropes framework itself, which is constituted of a three-layer categorization (micro, meso, and meta) structure. Second, based on coding resulting from these trope categorizations, we present a grammar of rumors supported by quantitative findings from the dataset we examined. Within these quantitative findings, the co-occurrence network of micro tropes (Figure~\ref{fig:network_community_coloring} \& Figure~\ref{fig:network_framework_coloring}) stands out as one of the most useful findings, demonstrating how clusters of micro-tropes visible in the data can be viewed as sub-narratives that drive rumors; sub-narratives that are each composed of different types of tropes that function as building blocks.

\subsection{Tropes Framework}

Once calibrated, three distinct categories emerged in the framework: micro-level tropes documenting the most-granual level of trope categories, meso-level tropes grouping similar micro tropes together based on their thematic forms and functions, and meta-level categories describing these building blocks by the role each category plays in a rumor. Meta-level categories communicate the function that the items of these categories fulfill in a narrative: actors, objects, actions taken by actors or on objects, scenes and settings, and suggested remedies. At the meso-level, the categories incorporate more detail and framing, for instance, the actor meta category includes meso tropes: “victims,” “villains,” and “heroes.” Under each meso-level trope, the micro-level tropes consist of the specific tropes that function as building blocks of election rumors—for example, the Democratic Party as an actor, pre-filled ballots as an object, or a polling center as a setting.

\subsubsection{Meta-Level Categories}

In our framework, we have six meta-level categories within which specific tropes fall: Actor, Actor-action, Object, Object-action, Scene and Setting, and Suggested Remedies. These meta categories are the skeleton of rumor narratives that structure the storyline and set the scene (e.g., actors, objects, actions). Within this framework, we do not code the plot (e.g., election fraud or cover-up being the fraud) since almost all of our rumors have the same underlying plots of either fraud or vulnerability to fraud. Therefore, the actor-object-action oriented framework covers the themes, recurring plotlines and motifs of the election rumors (e.g., hacked machines, destroyed mail-in ballots) rather than focusing on the underlying plot itself (e.g., election fraud). 

\begin{table}[htbp]
\centering
\begin{tabular}{|>{\raggedright\arraybackslash}p{2.5cm}|>{\raggedright\arraybackslash}p{6cm}|>{\raggedright\arraybackslash}p{4.5cm}|}\hline
\textbf{Meta categories} & \textbf{What it captures} & \textbf{Typical rumor question} \\
\hline
Actors & Social roles for those involved in rumors, and the moral valence assigned to their actions & Who is blamed, harmed, or fighting back? \\
\hline
Actor-Action & Alleged acts by actors, often implied as intentional & What did the actor allegedly do? \\
\hline
Objects & Entities around which rumors form & What object becomes evidence? \\
\hline
Object-Action & Actions taken on or with the central objects of rumors & What happened to the object in a rumor? \\
\hline
Scene/Setting & Frequently named times and places that `set the stage' for rumors & Where/when does the rumor gain suspicion? \\
\hline
Remedies & Mobilization or policy demand which result from or are justified by rumors & What response does the rumor demand? \\
\hline
\end{tabular}
\caption{Meta-level categories and their functions in a rumor}
\Description{A three columned, seven rowed table (including header row). The columns are the meta categories, what each category captures, and the typical rumor question the meta-category helps answer.}
\label{tab:rumor_categories}
\end{table}

To demonstrate how these categories are formed and interact with each other, consider the following example summary from our data (a rumor from 2020 called “Digital dumps: Michigan 128,000 votes”):

\begin{quote}
\textit{The tweets revolve around widespread allegations of voter fraud and election irregularities during the 2020 U.S. presidential election, particularly focusing on \textbf{Michigan}. A central claim is that approximately 130,000 to 138,000 ballots were mysteriously \textbf{"found"} or \textbf{"dumped" late at night} in Michigan, with all or nearly all votes going to \textbf{Joe Biden} and none to \textbf{Donald Trump}, which is described as \textbf{statistically impossible} and suspicious by many users. There are specific narratives alleging that these ballots were delivered in a \textbf{convoy of vehicles} including a truck, a Chrysler 300C, and a Ferrari around \textbf{4 a.m.}, and that \textbf{Republican poll watchers were blocked} from observing the counting process in Wayne County, Michigan. The tweets accuse \textbf{Democrats, election officials, and media organizations} of conspiring to steal the election through these alleged fraudulent activities, with some also criticizing \textbf{Big Tech platforms} like Twitter for \textbf{censoring} related content. Additionally, former President Donald Trump and his campaign are mentioned as \textbf{challenging the election results legally}, claiming victories in key states such as \textbf{Pennsylvania, Georgia, North Carolina, and Michigan} despite ongoing counts and court dismissals of their lawsuits due to lack of evidence. The overall narrative portrays a deep distrust in the electoral process, calls for \textbf{investigations}, and demands for \textbf{recounts} or \textbf{legal challenges}, while some tweets mock or refute these claims, emphasizing the lack of credible evidence.}
\end{quote}

The above example contains examples from every meta-category in our framework (we bolded the elements that led to specific codes to make it easier for readers to parse). First is the Scene and Setting: the rumoring revolves around Michigan, a swing state, and allegations about events taking place late at night. Second, the Actors are numerous, with Joe Biden, Democrats, Media, and Big Tech named as villains, while Donald Trump and his campaign are both victims of alleged fraud and heroes fighting back through litigation. Third, the Objects include the unofficial “convoy” of vehicles allegedly carrying ballots at 4am and the supposedly “statistically impossible” results. Fourth, Actor-Action focuses on alleged censorship by Media and Big Tech, who are seen as working in tandem with alleged Democrat fraudsters. Fifth, Object-Action focuses on the “found” ballots as the alleged mechanism through which fraud occurred. Here, the ballots are not literally found, but are referred to sarcastically by those who claim Democrats “find” ballots whenever they are losing a race. Last are the Suggested Remedies, including investigations, recounts, and lawsuits. Each category functions as a piece of the puzzle, coming together to form the rumor and add context to the others.

\stepcounter{footnote}
\begin{figure}[H]
    \centering
    \textbf{The Three-Layer Nested Structure of the Tropes Framework}\\[0.5em]
    \includegraphics[width=0.8\textwidth]{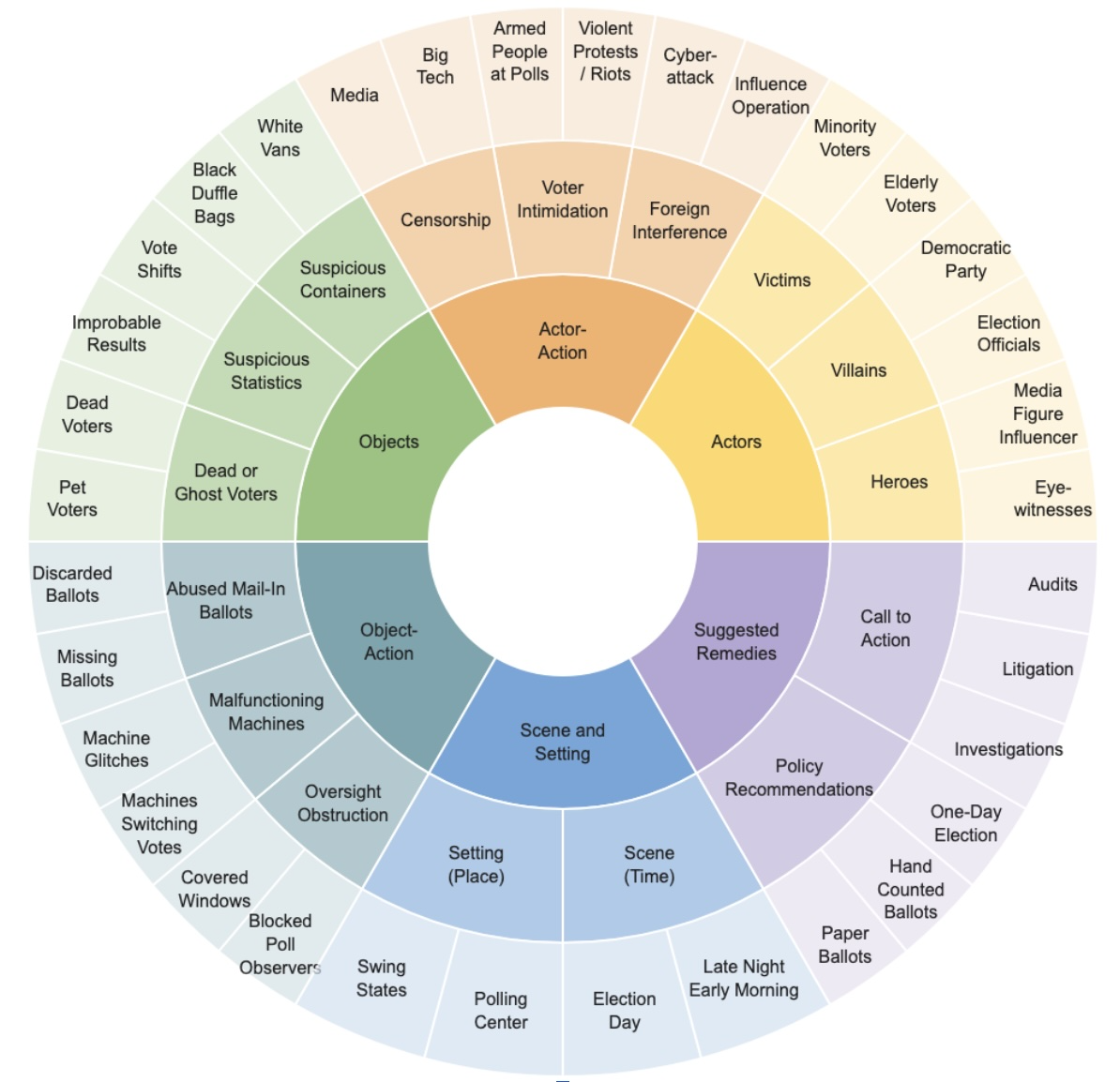}
    \caption[The three-layer nested structure of the tropes framework]{This wheel chart is a partial representation of the election rumor tropes framework. All meta categories are shown, while selected meso categories and micro tropes are displayed as examples. The innermost circle shows meta categories, such as Objects or Actors; the middle circle shows meso categories, such as Suspicious Statistics or Villains; and the outer circle shows micro tropes, such as Improbable Results or Election Officials.\protect\footnotemark[\value{footnote}]}
    \Description{A circular wheel chart with three concentric rings. The inner ring lists meta categories of election rumor tropes, the middle ring lists example meso categories within each meta category, and the outer ring lists example micro tropes within each meso category.}
    \label{fig:wheel_chart}
\end{figure}
\footnotetext[\value{footnote}]{The wheel chart is selective below the meta level. For Scene and Setting and Suggested Remedies, all meso categories are included, but not all micro tropes are displayed. For all other meta categories, both meso categories and micro tropes are selectively displayed rather than shown in full.}

In short, meta categories explain the main structure of the rumors that we analyzed at the highest level of abstraction (Actors, Objects, Actions, Scenes \& Settings, and Remedies), meso categories describe the type of actors (e.g. heroes and villains) and objects (e.g. mail-in ballots or voting machines), and micro categories detail the actual tropes in the most granular level (e.g. eyewitnesses as heroes, discarded mail-in ballots, hacked machines). For a full list of the micro- and meso-tropes, we have included the full coding scheme in Appendix A.

Figure~\ref{fig:wheel_chart} presents a partial illustration of the framework’s three-layer nested structure. Meta-categories (e.g., Actor-Action) are presented in the innermost circle, meso tropes (e.g., voter intimidation) in the middle circle, and micro tropes (e.g., armed people at the polls) in the outermost circle. Throughout the remainder of the paper, Figure~\ref{fig:wheel_chart} can serve as a guide for understanding the nested structure and navigating what is otherwise a fairly complex framework.

\textbf{Actors}\\
The Actors meta-category captures the people, groups, and institutions through which election rumors assign responsibility and victimhood. Even though the details of each rumor are shaped by the events of the day, the types of actors that appear in them often repeat: some actors are portrayed as committing fraud (villains), some as exposing it (heroes), and some as suffering from it (victims). In this sense, the actor category adds agency and moral valence to the story’s characters. A specific actor may change roles across different rumors, but the trope structure remains intact, as people fit new rumors into familiar buckets of villain, hero, and victim.

Our framework codes more than 20 actors, some of whom can appear in multiple narrative roles depending on the rumor. The Democratic and Republican parties, for example, may be portrayed as heroes fighting fraud, villains committing fraud, or victims of the other side’s alleged fraud. Other actors are far more role-consistent across rumor narratives. Elderly voters, whose ballots are often framed as being harvested or manipulated, are almost always cast as victims. Eyewitnesses, by contrast, are typically presented as heroes who reveal hidden fraud or cover-ups, whereas election officials are more often depicted as villains accused of enabling those cover-ups or carrying out fraudulent activities.

\textbf{Actor-action}\\
Actor-action tropes revolve around an actor’s activities. Our framework contains 30 micro tropes under eleven actor-action meso categories. Foreign interference (44 occurrences, 11\% of group) and voter intimidation (78 occurrences, 19\% of group) are some of the most recurring examples within the actor-action umbrella. For example, foreign interference includes micro tropes such as influence operations, machine manipulation or cyberattacks on election infrastructure, and manipulated ballots. These micro tropes often co-occur with designated actors such as Russia, China, Iran, or Venezuela. Within the voter intimidation meso category, we have micro tropes such as armed people at the polls, police intimidating voters, and anonymous intimidating calls or texts meant to deter certain voters from voting. In each meso category, the micro tropes capture the specific mechanisms through which fraud allegedly occurs. They explain, for instance, how the Chinese government is alleged to have interfered with U.S. elections through online influence campaigns or cyberattacks, or how intimidation is said to occur through Antifa or Proud Boys protesting at voting locations to deter people from voting.

\textbf{Objects}\\
The Objects meta-category captures tropes in which the central element is a material object, such as containers, vehicles, ballots, or envelopes. We identified over 25 object tropes, nested under seven meso categories. One illustrative example is the “white vans” trope. Although the specific allegations differ across cases, rumors involving white vans consistently portray them as suspicious transporters facilitating fraud, whether by delivering fake ballots, stealing valid ballots, or transporting non-citizens to polling places. Other object-centered tropes focus on the ballot or envelope itself, interpreting a design feature, irregularity, or flaw as evidence of manipulation. For instance, Sharpies that bled through ballots were framed by some rumors as a deliberate effort to invalidate Republican votes in Arizona during the 2020 presidential election \cite{prochaskaMobilizingManufacturedReality2023b}. Likewise, ballot layout decisions, candidate ordering, or misspellings may become central to rumors that treat ordinary design choices or errors as intentional signs of fraud.

\textbf{Object - Action}\\
Object-action tropes, on the other hand, capture elements of rumors where the fraudulent action is happening to or through an object. Our framework provides nine meso categories and 29 micro tropes within the object-action umbrella. Abused mail-in ballots and malfunctioning machines are particularly common meso categories within this meta-category. In abused mail-in ballots, as in our example of USPS dumping mail-in ballots, the object-action centers on mail-in ballots being discarded. This meso category also includes micro tropes such as destroyed, stolen, lost or found/discovered, and missing, which capture the wide variety of rumor tropes focused on abused mail-in ballots. In this case, the distinction between these categories is blurry, as ballots being lost or missing may be used interchangeably, and dumped, destroyed, or discarded may mean the same thing or carry slightly different meanings depending on the rumor incident. We retain some of these seemingly overlapping codes because we aimed to match the language of the codebook as closely as possible to the language used in the rumors themselves. By contrast, within malfunctioning machines, we observe more distinct tropes, including machine “glitches,” tabulators or machines being down, and machines flipping or switching votes from one candidate to another. 

The categories of objects, object-action, actors, and actor-action are conceptually adjacent, but are separated based off of the primary role the trope plays in the rumor narrative. For tropes in the objects category, the object itself is cast as evidence of suspicion or fraud. For example, a white van may be presented as suspicious simply by appearing in the story, regardless of whether a specific action is described. Similarly, in the actors category, the actor is central as a hero, villain, or victim, regardless of their particular action. The actor as a character and agent is the key, recurring trope. By contrast, the object-action and actor-action categories focus on the specific actions done with/by the object or actor. In the object-action category, the object becomes suspicious through an alleged action involving it — for example, electronic voting machines, for instance, are typically framed as suspicious not merely because they exist, but because they are said to flip votes or are vulnerable to hacking. As we describe in the Methods section, we structured the framework this way in part to account for discourse on social media, and in part to account for the storytelling dynamics that drove the rumoring we observed.

Similarly, for tropes in the actor-action category, the trope is tied more directly to what the actor is alleged to do. In other words, the rumor is organized not simply around who the actor is, but around the action that makes that actor suspicious or consequential, such as foreign interference, voter intimidation, or censorship. It is important to separate action-based tropes that are occurring through the presence of particular objects, and those by particular actors, which is why the codebook maintains these two categories rather than a more ambiguous, hypothetical “action” category.

\textbf{Scene and Setting}\\
The meta-category of scene and setting captures tropes that define the location, timing, or broader context in which alleged fraud or suspicious activity is said to occur within a rumor instance. Our framework includes nine micro tropes under two meso categories. Across these rumors, scene and setting tropes help structure where and when suspicious activity is claimed to happen. One of the most recurring tropes not only within this category but across our broader coding scheme is the framing of swing states as especially suspicious settings for fraud. These rumors often focus on close contests in electorally pivotal states, where relatively small shifts in vote totals could determine the winner. In this sense, swing states are cast not simply as important electoral battlegrounds, but as settings in which fraud is claimed to be most strategic, frequent and consequential. Another common scene-setting trope, often connected to the swing state trope, is the Democrat-run urban center trope. In these right-leaning rumors, a Democratic stronghold within a swing state is portrayed as the epicenter of fraud. These rumors often imply that Democratic control of urban centers allows local election officials and poll workers to coordinate irregularities or fraudulent activities.

\textbf{Suggested Remedies}\\
Finally, the meta-category of suggested remedies captures the ways rumor instances move beyond diagnosis and propose responses to alleged fraud or suspicious activities. We identify sixteen micro tropes within two main forms (meso-level categories): policy recommendations and calls to action. Policy recommendations are typically tied to the specific mechanism of fraud alleged in the rumor. For example, rumors about voting machines flipping votes often lead demands to abolish machine voting and return to paper ballots and hand counting. Rumors portraying mail voting as vulnerable to manipulation often produce calls to eliminate mail voting in favor of in-person voting and single-day elections.

Calls to action differ in that they seek immediate behavioral or institutional responses rather than future policy reform. In rumors centered on mail-ballot fraud, participants often urge others to vote in person in order to avoid perceived risks of ballot disposal, theft, or invalidation. More broadly, these calls may include demands for audits, recounts, and investigations into alleged irregularities, or litigation challenging election processes or results. Some rumors also encourage collective mobilization, including protests or armed poll-watching justified as protection against intimidation or fraud. Overall, suggested remedies appear in a large share of the rumor incidents we examined (70.8\%), making them one of the most common features of election-rumor narratives.

\subsection{The Grammar of Election Rumors: Rumor Building Blocks}
The framework we introduce above provides a structure for understanding the primary elements present in the election rumors we analyzed. However, the tropes we document do not occur in isolation. Instead, they often clustered together, combining to tell the stories that made up and propelled the rumors we examined. On its own, no single trope carries enough significance or meaning to drive a rumor to prominence. But rumors combine different tropes to tell variations of stories that carry with them both the novelty of emergent events and the familiarity of tropes that have been used time and time again.

Using our framework, we analyzed which tropes occurred most frequently in general as well as which tropes co-occurred most frequently together. By examining which tropes co-occurred, we were able to see what patterns and storylines were present across rumors. The couplings of different actors, objects, actions, settings, etc. allow us to see both central narratives and important, but less common relationships that were present in relatively fewer rumors. 

Emerging from this analysis is what we see as the second-most important contribution of our work beyond the framework itself: a grammar of election rumors. We find that the construction of rumors we examine follow a set of rules somewhat similar to the grammatical rules guiding the construction of legible phrases and sentences. Within the rules of grammar, different types of words (and word forms) are combined in specific ways to communicate different types of meaning. For example, every sentence must, at a minimum, contain a subject and a verb. Together, they describe a thing, and what that thing does or is doing. 

Rumors are of course more complex than a single sentence, so the grammar we propose is necessarily more complicated than such a simplistic example. Based on the analysis summarized below, we propose that election rumors follow specific “grammatical” rules. Within this grammar, most election rumors contain an actor, an object, and an action. Additionally, the majority of rumors in the 2020 \& 2022 U.S. Election Rumor Dataset also contained a Scene \& Setting or a Suggested Remedy that added additional texture to any given rumor. Those categories are not enough on their own to make a rumor, but they are frequently included in ways that increase the salience and potential utility of the rumors in question. In the following sections, we discuss how the micro- and meso-tropes come together to form the Meta-Categories that form this grammar of election rumors.

%which we represent in the following expression:

%$R = A \land (O \lor OA) \land (AA \lor OA) \land [SS \lor SR]$

%In plain language, this can be understood as saying an election Rumor (R) is composed of an Actor (A), necessarily an Object (O) or Object-Action (OA), necessarily an Actor-Action (AA) or Object-Action (OA), and optionally a Scene \& Setting (SS) or Suggested Remedy (SR). 

\subsubsection{Grammar is Based on a Repeated Core: Meta-Category Frequency}

\begin{figure}[!htbp]
    \centering
    \includegraphics[width=.9\textwidth]{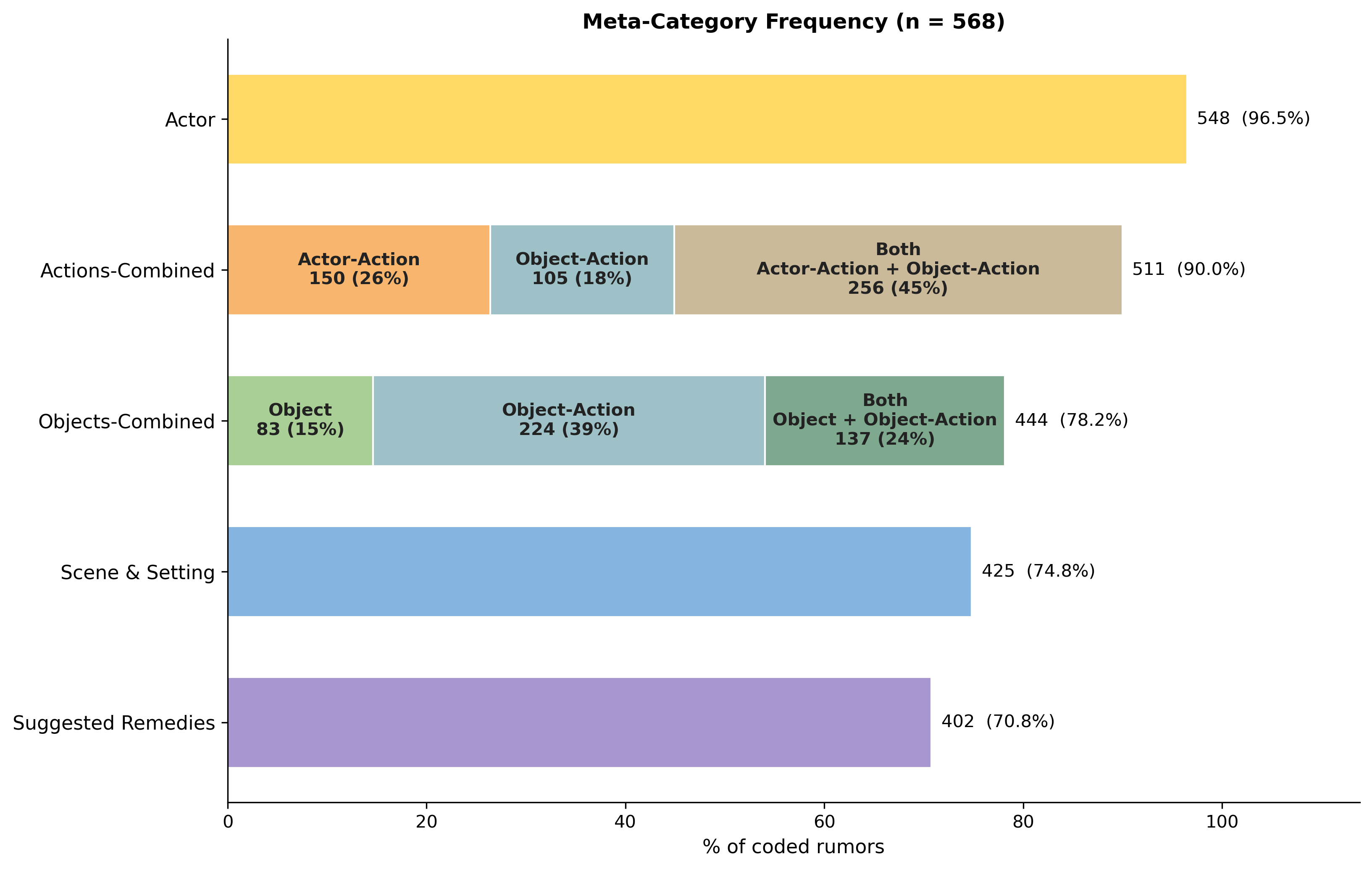}
    \caption{Frequency of Meta-Categories across coded rumors.}
    \Description{A barchart showing the percentage of each Meta-category present across rumors. Actor (96.5\%), Actions-Combined (90\%), Objects-Combined (78.2\%), Scene \& Setting (74.8\%), Suggested Remedies (70.8\%).}
    \label{fig:metacategory_barchart}
\end{figure}

We first visualize our Meta-Categories to show a high-level view of the major rumor building blocks that form the grammar we describe. Figure~\ref{fig:metacategory_barchart} shows the relative frequencies of each of these categories across the 2020 \& 2022 U.S. Election Rumor Dataset. The Actor category showed up in over 96\% of the rumors we coded, supporting the interpretation that actors are a core, consistent piece of election rumors. Although they are separate in our framework, we group Objects and Object-Action into a single category (“Objects-Combined”) for this part of the analysis. Similarly, we group the action categories in “Actions-Combined” in an effort to group elements of the framework based on their relationship to the rumors as a whole. 

As we discussed in the Methods section, Object-Action as a category captures many objects that are indelibly intertwined with the actions they are associated with. Because of this, many of the codes in the Object-Action category describe objects that do not exist in the Objects category, motivating our decision to group the two meta-categories into a single “Objects-Combined” category for the quantitative-descriptive analysis. Therefore, when this composite category is present we know that a major component of a rumor involves objects whose meaning is informed by the other “grammatical” components  of the rumor. Similarly, the “Actions-Combined” composite category captures the core actions that a particular rumor revolves around.

Figure~\ref{fig:metacategory_barchart}, therefore, demonstrates that the election rumor grammar we propose in that Actors, Objects (composite), and Actions (composite) are the most frequently occurring meta-categories. Moreover, Scene \& Setting, although optional in our grammar, is clearly a salient dimension of election rumoring in our dataset, being present in almost 75\% of the rumors we examined. Similarly, Suggested Remedies was present in roughly 70\% of the rumors we coded, suggesting that many of the rumors in our data were not simply descriptive, and were instead intended to motivate action of some kind, aligning with prior work (e.g. \cite{prochaskaMobilizingManufacturedReality2023b}). 

To further validate our proposed grammar, we examined how many rumors from the 2020 \& 2022 U.S. Election Rumor Dataset conform to the expected structure. Table~\ref{tab:category_presence_by_grammar} shows that approximately 70\% of the rumors align with the grammar while about 30\% are missing one of the central elements. Amongst those with “proper” grammar, Scene \& Setting or Suggested Remedies are also present in 375 of the 397 rumors (~94\% of rumors with proper grammar), further supporting the interpretation that although they may not be a core element of every rumor, they are important features of election rumors. Within the grammar analogy, we view these categories similar to adjectives that add depth and nuance to a rumor, but are insufficient on their own to carry enough meaning to drive a rumor.

It is important to understand how rumors with “improper” grammar may be different from those with the expected structure. Of those with improper grammar, by far the most common missing element was the Objects-Combined composite group, with 99 rumors (~58\% of rumors with missing elements) missing an object. In reviewing rumors missing an Objects-Combined code, we saw a range of reasons this code was missing. In some cases, rumors such as the “voting info: Too Late to Send Mail Ballots” rumor, involved objects peripherally (in this case mail-in ballots) but they were not discussed in a way that led to them being coded. We do not have a generic code for capturing when “mail-in ballots,” for example, are mentioned because they are mentioned in so many contexts and so often that the code would be applied to most rumors. In the case of the “voting info: Too Late to Send Mail Ballots” rumor, the claims focused more on the USPS as an actor than on the ballots themselves. The second type of rumor that was missing the Objects-Combined category focused primarily on actors. For example, the “intimidation: NC BLM” rumor revolved around claims that Black Lives Matter (BLM) and “Antifa” activists were intimidating voters at polling centers in North Carolina. Here, objects such as ballots take a back seat to the central claims of voter intimidation that drove the rumoring. In both of these cases it's less that objects don’t exist or aren’t relevant, and more that they aren’t central to the rumor in question.

% Table 1: optional elements x grammar group
\begin{table}[htbp]
\centering
\begin{tabular}{lrrrr}
\toprule
 & \multicolumn{2}{c}{Proper grammar} & \multicolumn{2}{c}{Improper grammar} \\
\cmidrule(lr){2-3} \cmidrule(lr){4-5}
Optional elements & $n$ & \% & $n$ & \% \\
\midrule
Both optionals & 246 & 62.0 & 65 & 38.0 \\
Scene \& Setting only & 67 & 16.9 & 47 & 27.5 \\
Suggested Remedies only & 62 & 15.6 & 29 & 17.0 \\
No optionals & 22 & 5.5 & 30 & 17.5 \\
\midrule
\textbf{Total} & \textbf{397} & \textbf{100.0} & \textbf{171} & \textbf{100.0} \\
\bottomrule
\end{tabular}
\caption{Optional elements present in rumors with proper vs.\ improper grammar). Proper: $n=397$ (69.9\% of rumors); improper: $n=171$ (30.1\%); total $N=568$.}
\label{tab:optionals_by_grammar}
\end{table}

% Table 2: meta-category presence by grammar group
\begin{table}[htbp]
\centering
\begin{tabular}{lrr|rrrrr}
\toprule
 & \multicolumn{2}{c|}{All ($n=568$)} & \multicolumn{2}{c}{Proper ($n=397, 70\%$)} & \multicolumn{2}{c}{Improper ($n=171, 30\%$)} & \\
\cmidrule(lr){2-3} \cmidrule(lr){4-5} \cmidrule(lr){6-7}
Meta category & $n$ & \% & $n$ & \% & $n$ & \% & Gap (pp) \\
\midrule
Object-Action & 361 & 63.6 & 347 & 87.4 & 14 & 8.2 & +79.2 \\
Object & 220 & 38.7 & 185 & 46.6 & 35 & 20.5 & +26.1 \\
Suggested Remedies & 402 & 70.8 & 308 & 77.6 & 94 & 55.0 & +22.6 \\
Actor-Action & 406 & 71.5 & 304 & 76.6 & 102 & 59.6 & +16.9 \\
Scene \& Setting & 425 & 74.8 & 313 & 78.8 & 112 & 65.5 & +13.3 \\
Actor & 548 & 96.5 & 397 & 100.0 & 151 & 88.3 & +11.7 \\
\bottomrule
\end{tabular}
\caption{Presence of each meta category across all rumors and within the proper- and improper-grammar groups. Gap is the percentage-point difference (proper $-$ improper).}
\label{tab:category_presence_by_grammar}
\end{table}

% Table 3: missing-pattern breakdown of improper-grammar rumors
\begin{table}[htbp]
\centering
\begin{tabular}{lrrr}
\toprule
Missing Meta-Category & $n$ & \% of improper & \% of all \\
\midrule
Missing Object group & 99 & 57.9 & 17.4 \\
Missing Action group & 31 & 18.1 & 5.5 \\
Missing Object group + Missing Action group & 21 & 12.3 & 3.7 \\
Missing Actor & 14 & 8.2 & 2.5 \\
Missing Actor + Missing Object group + Missing Action group & 3 & 1.8 & 0.5 \\
Missing Actor + Missing Action group & 2 & 1.2 & 0.4 \\
Missing Actor + Missing Object group & 1 & 0.6 & 0.2 \\
\bottomrule
\end{tabular}
\caption{Combinations of required grammatical components missing from improper-grammar rumors; $n=171$ improper of $N=568$ total).}
\label{tab:missing_patterns}
\end{table}

Similarly, the Actions-Combined category is the second most often missing category, missing from 31 rumors. Like the Objects-Combined category, examining the rumors missing a coded action reveals that more often than not actions were implied or assumed, but were not central to the rumor in question. For example, the “Ineligible Voters: Illegal Immigrant Registrations” rumor focused on claims that non-citizen voters were illegally registered to vote. The rumor itself did not focus on the act of registration (there were some references to automatic voter registration when getting a drivers license, but we viewed this as passive and therefore not an action in our coding scheme) or on the act of voting, instead focusing more on the claims that voter rolls were “dirty.” In this case the actions of registering or voting are implied but not the focal point of the rumor. Another example is the “Statistics: Benford's Law” rumor which is similar to a class of rumors that focuses on data visualizations and statistics as signals of alleged fraud. In cases like this, the rumoring tended to focus more on the statistics as a signal of some, often nebulously defined, form of fraud. We had no generic “fraud” code for these cases, resulting in them being coded without an action.

\subsection{Micro-Trope Co-Occurence: A Central Story with Familiar Sub-Narratives}

The most granular view of our findings involves examining the relationships between micro-tropes to better understand how they come together as part of the grammar we have outlined above. However, a major challenge is simply the sheer number of micro-tropes we documented alongside the fact that every micro-trope has a different relationship with each of the other micro-tropes in our data. In order to better understand these multi-dimensional relationships, we generated a network that shows how these tropes tend to occur together.

\begin{figure}[htbp]
    \centering
    \includegraphics[width=0.8\textwidth]{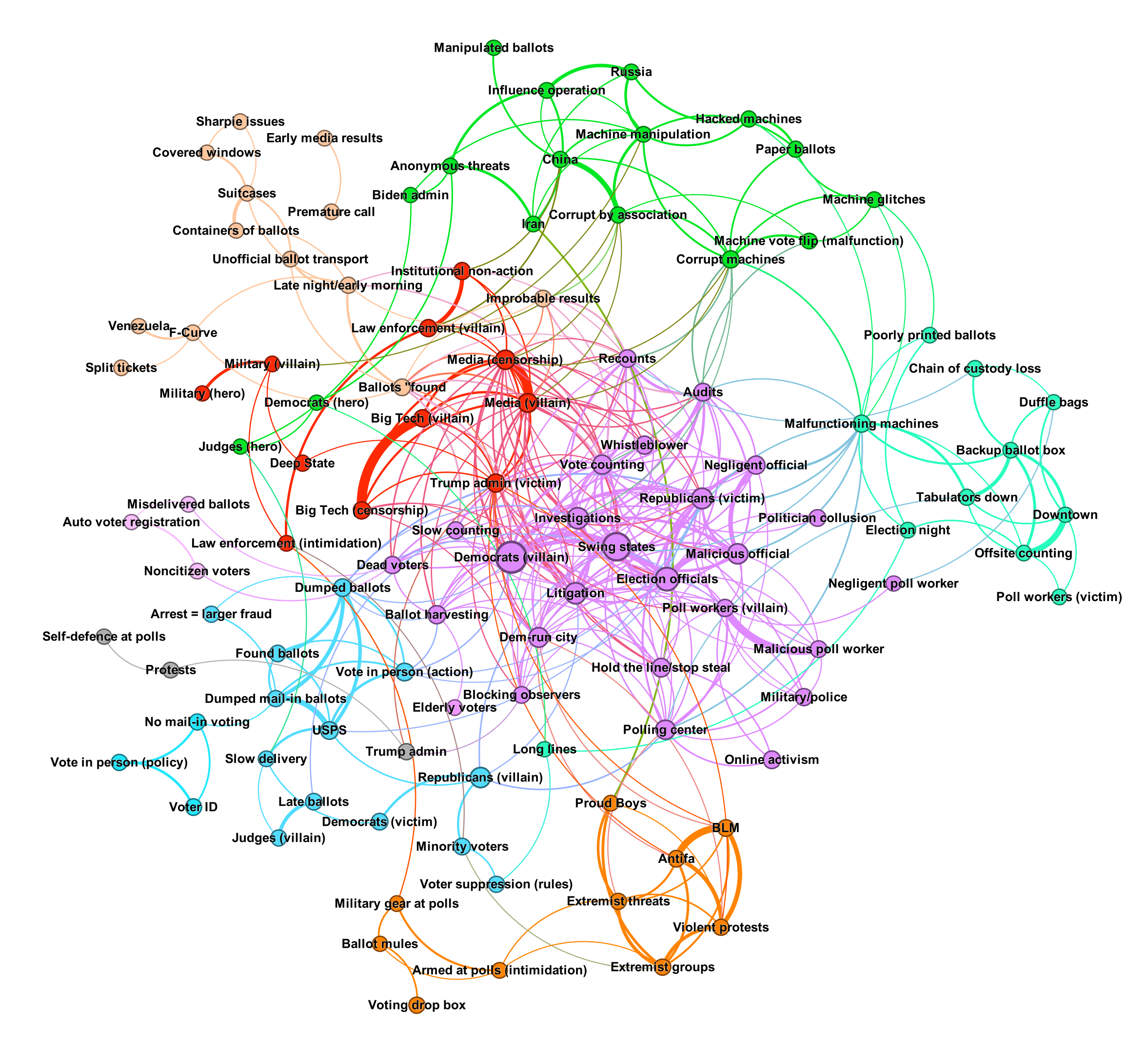}
    \caption{Network of co-occurring tropes colored using the Louvain clustering algorithm for community detection \cite{blondelFastUnfoldingCommunities2008}. Larger nodes indicate more frequently occurring codes and node size is capped to prevent some nodes from obscuring others. Code names were shortened to fit onto the graph, for a key connecting original codes to shortened labels, see Appendix B.
}
    \label{fig:network_community_coloring}
\end{figure}

Figure~\ref{fig:network_community_coloring} shows the most commonly co-occurring tropes across our dataset, colored by trope “community” using the Louvain clustering algorithm \cite{blondelFastUnfoldingCommunities2008}. Each node represents a micro-trope and each edge represents tropes that co-occur together over a Jaccard value of .1. We used Jaccard similarity to calculate the edges between nodes, which, for each code pair, is calculated as the number of rumors where both codes appear divided by the number of rumors where either code appears $J(A, B) = \frac{|A \cap B|}{|A \cup B|}$. Nodes are only connected to another node if they are above a Jaccard threshold of .1, and edges are weighted more heavily for higher values (which indicate higher co-occurence).\footnote{Functionally, this reduces the visibility of very common codes that co-occured with almost every other code (e.g. the “Democrats (villain)” code) while increasing the visibility of codes that occurred relatively infrequently, but when they were present they were often paired with the same other codes.} Two tropes coded within the same rumor appeared in the same broader discussion, but not necessarily in the same post or among people expressing the same viewpoint.

Figure~\ref{fig:network_community_coloring} shows several different clusters of tropes, each of which come together to support different, but related, rumor narratives. The central, most prominent cluster is visible in purple, and represents some of the most frequently occurring tropes across rumors. In our analysis, these tropes created a common foundation upon which many rumors are built. Within that foundation, Democrats are viewed as the primary villains who are often in charge of election administration in major, Democratic-leaning urban centers, such as Detroit and Atlanta. Rumours often falsely viewed them as exerting corrupt influence via election officials and poll workers alongside tactics such as ballot harvesting and fabricating ballots using the names of dead, albeit still registered, voters. Discourse was particularly prominent surrounding events in swing states, where Republicans were the most commonly occurring victims. Following the grammar of rumors, these tropes come together to suggest a problem definition from which proposed remedies emerge: calls for audits, recounts, investigations, and litigation alongside more nebulous calls to “stop the steal” or “hold the line.” 

Surrounding the central, purple cluster of co-occurring tropes there are numerous other clusters highlighted, each of which roughly corresponds to a frequently occurring subnarrative that different rumors revolved around. The red cluster (on the left of the purple center) revolves around distrust of media and “big tech” (largely Facebook and pre-Musk Twitter), who were often viewed as an extension of corrupt Democrats or their alleged handlers (the “deep state”). In addition to media organizations themselves being targets, numerous rumors were present that focused on media reporting that provided a signal that acted as “evidence” for allegations toward “mainstream media”. For example, rumors often discussed how the reported statistics displayed by the media were evidence of fraud (“Improbable results” trope) or how results were too slow to come in, signaling deception and allowing Democrats to “find” ballots to artificially inflate vote counts (“Slow counting” and “Ballots ‘found’” tropes). 

Similarly, the light green cluster (at the top center) revolves around voting machine and tabulator tropes, with a focus on corruption and foreign influence as vectors of machine manipulation. The teal cluster focuses on false allegations of corrupt offsite counting resulting from allegedly intentional machine errors, while the orange cluster primarily revolves around claims of voter intimidation or suppression by extremists and other groups. In contrast, the blue cluster (down and to the right, surrounding the purple cluster) represents rumor tropes that primarily focus on claims of mail-in voting fraud and/or voter suppression related to mail-in voting. The final sizable cluster (brown) focuses on suspicious signals related to allegations of fraudulently inserted ballots, either through suspicious spikes in ballots counts, early media calls, or the physical injection of ballots via suspicious containers late at night.

It is worth noting that the Louvain community detection algorithm we used is, like all methods, imperfect. In reality, many of the identified clusters have fuzzier boundaries than the color categories would suggest, in that many codes in one cluster co-occurred with codes present in other clusters. However, it is helpful to visualize these relatively distinct clusters to see where major convergences are, which resulted in the loosely defined subnarratives we summarize above. These subnarratives align with our higher level thematic analysis as well, which provides a second layer of validation of the clusters even if they are not quite as distinct as they appear in the visual. 

\begin{figure}[htbp]
    \centering
    \includegraphics[width=0.8\textwidth]{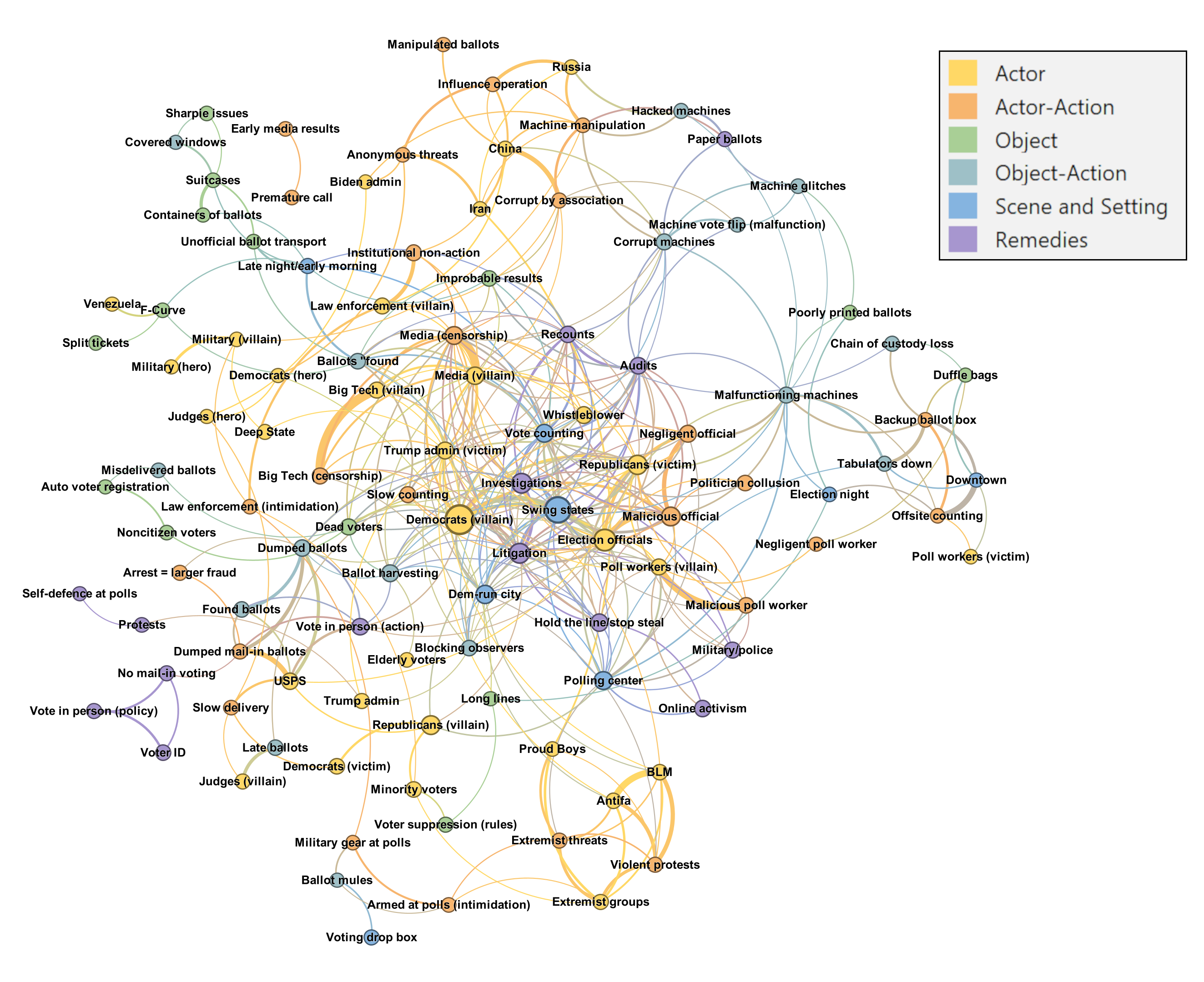}
    \caption{Network of co-occurring tropes colored based on Meta-Level Categories. }
    \label{fig:network_framework_coloring}
\end{figure}

To complement this view of the network of co-occurring codes, we also colored the network based on the Meta-Level Category to which each micro trope belongs, which is visible in Figure~\ref{fig:network_framework_coloring}. This is a strong piece of evidence supporting our interpretation of tropes as the “building blocks” of election rumors - building blocks that come together to form the grammar we propose. Figure~\ref{fig:network_framework_coloring} shows how tropes often appear alongside tropes from different Meta-Level Categories, collectively constructing narratives and subnarratives that rumors are made up of. For example, the central cluster consists of tropes from the Actors, Actor-Action, Object-Action, Scene \& Setting, and Suggested Remedies categories, with each trope providing a piece of the larger story. Similar relationships can be seen throughout the graph, highlighting how rare it is for a trope category to appear in isolation. 

In the data we analyzed, not every rumor contained tropes from every Meta-Level Category. Instead, rumors were composed of different combinations that we argue to be the grammar of rumors, tying novel events to familiar tropes that fit within larger narratives of election fraud within this grammar. One of the most salient findings highlighted in Figures~\ref{fig:network_community_coloring} and \ref{fig:network_framework_coloring} is that the central cluster is made up of tropes from every meta-category. This resonates with prior work documenting the centrality of characters, time, and place to a story. If a reader focuses on the central cluster, an at-a-glance read of the network provides the most central information an unfamiliar reader might want to know: who the central antagonists and protagonists are, when and where actions are alleged to take place, what the actions are, and what the proposed solutions are. 
Although the rest of the network is immensely useful and relevant, that core cluster provides an immense amount of information in a very small space in a way that represents the core narratives driving rumoring in our data. A slightly deeper read suggests that, in electoral rumoring, the “suggested remedies” category can be seen as the “moral” of the story from a narrative perspective: the takeaway readers are left with at the end of the rumor, or the action the story implies should follow. From a political strategy perspective, these suggested remedies also point to the goals the rumor is trying to advance. It is not an accident that so many influencers and political elites focus on swing states, and that the suggested remedies consist of investigations, audits, and other pathways to influencing an electoral outcome, including longer-term strategies such as implementing voter ID laws or abolishing voting machines or mail-in voting.

\section{Discussion}
\subsection{Examine the Past to Make Sense of the Future}
The framework and grammar we introduce in this paper has utility for unpacking, analyzing, and anticipating election rumors. It can also be adapted to rumors outside the election context by using the meta-categories and its grammar to structure rumor analysis and applying the multi-dimensional coding scheme to new domains. One of the most direct use cases for our framework, however, is for researchers examining rumors about U.S. elections, both historically and in current and future electoral contexts. Although the data we examine in this study comes from rumors surrounding the 2020 and 2022 U.S. elections, we argue that many of the meso- and micro-tropes are likely to remain relevant and reappear in future elections. Our nested tropes framework therefore has continued utility for current and near-future elections, while also remaining flexible enough to be expanded when new tropes emerge.

This paper makes three main contributions. First, it shows that election rumors, which often appear unpredictable or random in their emergence and circulation, can be systematically unpacked, analyzed, and shown to follow recurring narrative patterns that follow a consistent grammar. Second, it provides researchers, practitioners, and the broader public with a framework for dissecting and diagnosing new rumors. By identifying a rumor’s building blocks, or tropes, the framework helps reveal familiar patterns and supports timely responses to real-time rumors based on what has appeared before. Third, the tropes framework, and especially the co-occurrence of micro tropes, helps us anticipate what kinds of events are likely to generate rumors and what kinds of actors, objects, actions, and settings are likely to be involved in crisis moments.

This relative consistency in election rumor grammar is potentially useful in numerous ways. Perhaps most directly, it offers substantial predictive utility that can inform the detection and rapid analysis of novel rumors that might appear new but are, in reality, derivations of older stories. For someone familiar with the last decade of U.S. election information space or the 2020 and 2022 U.S. Election Rumor Dataset, and with how ballots are typically counted and reported, it becomes immediately apparent for instance that a data visualization showing intermittent spikes in vote counts is unlikely on its own to constitute evidence of anything suspicious. Future work could explore adapting this framework to enable human-in-the-loop automated rumor detection, as well as innovative, interactive, and gamified approaches to communicating about recurring election rumors.

In this vein, the framework has immense utility for supporting and structuring traditional research surrounding election rumors and elections more generally. By structuring the framework in a way that accounts not just for the topics discussed, but their functions and roles relative to rumors and narratives, the framework can act as a starting point for analyzing both historical and emergent rumors. For example, analyses gauging relative virality of specific types of tropes and their co-occurring neighbors may help guide investigations into questions of causality or other more narrowly tailored hypothesis testing. More generally, our identification of the grammar of rumors can inform examination in other contexts. Future research could explore what the salient actors, objects, actions, etc. are within rumors around online health advice or rumors emerging after extreme weather events, and whether there is a consistent grammar across types of rumors or if each domain has a different set of rules and patterns.

\subsection{Informing Election Communication}
Increasingly, election officials and others are being asked to communicate nuanced election processes which were previously not of interest to the public. Election infrastructures, processes, and procedures also vary by state and are often quite complex, making it hard to engage in “prebunking” simply because it can be difficult to know which aspects are important to communicate. We propose our framework as a first step for guiding communication towards the objects, actors, and actions that are the most common subjects of rumoring. Moreover, with some adaptation and interpretation, election officials seeking to maintain electoral integrity may be able to use the framework to inform threat assessments surrounding challenges that may dramatically impact how they administer elections. 	
While tropes are often recurrent, the online attention economy and news cycles result in an ebb and flow in which kinds of tropes are salient at a given moment. For example, while claims of noncitizen voting existed in 2020, they were few and far between and accounted for only a negligible portion of the data relative to our full dataset. By contrast, in 2024, claims of noncitizen voting were a major driver of election rumors and related propaganda \cite{andermanDamageConspiracyTheories2024}. The 2026 and 2028 election cycles are likely to have their own focuses, with some tropes becoming ascendant in the public eye while others fade into the background until they once again become salient to audiences.
Alongside this ebb and flow, new tropes may emerge, and old tropes may acquire additional meanings. In 2026, for example, we anticipate rumors surrounding ICE (Immigration and Customs Enforcement) being present at polling locations to intimidate voters. This type of rumor is present in our historical data, but in 2026 it carries additional weight—and plausibility—due to the extralegal and violent actions ICE has taken since 2024 (e.g. \cite{rubinICEAgentWho2026}).
To ensure that the framework retains its utility for researchers and communicators, we intend to update it with novel tropes as they emerge in future election cycles. We encourage others to use the provided framework in a similar way: with Meta and Meso categories providing structure, while the Micro tropes constitute a set of known examples rather than a complete list.

\section{Conclusion}
As with any study, our work has important limitations that should be taken into account when interpreting our findings and the framework itself. First and foremost, we developed the framework based primarily on English-language Twitter data focused on the administration of U.S. election processes. This is a very specific cultural, political and electoral context, and findings of specific election administration rumor tropes may not transfer unaltered to settings outside of the United States context. Events over the past five years have seen multiple global leaders attempt to make claims of election fraud \cite{bastosInsurrectionistPlaybookJair2023, leePoliticalDiscourseTwitter2024}, so there is precedence for similarities, but given the differences in electoral systems, it is unlikely that all of the micro-tropes transfer outside of a U.S. context. 

Additionally, the tropes we identify were those most common at the time on Twitter (now X). It is plausible that other platforms may have different norms, affordances, and audiences that incentivize the amplification of different tropes with which we are less familiar. We therefore urge caution in interpreting our framework as comprehensive, instead viewing these results as capturing a meaningful proportion of electoral rumoring tropes. For instance, our tropes are based primarily on written communication, whereas on more visually oriented platforms, tropes may instead be communicated through visual cues and recurring visuals (of tropes).

Similarly, our framework is primarily based on two election cycles: one presidential election and one midterm election. As we referenced earlier in the discussion, we have seen many of the tropes we document in multiple other election cycles operating at various levels of the U.S. election system (local, state, and federal races), but without access to the same systematically collected, comprehensive data it was impossible to replicate the quality of data available from 2020 and 2022. This means that there are likely nuances and low-virality tropes from other elections that may not be included in our framework or analysis.

Methodologically, we complemented traditional qualitative coding and interpretation with LLM-generated rumor summaries. This enabled us to evaluate a much larger dataset than would otherwise have been possible, but the summaries were necessarily limited in scope and tended to emphasize the most prominent themes, actors, and calls to action. Known issues such as hallucinations could also affect interpretation, although we removed summaries that appeared implausible based on our knowledge of the data. Additionally, we informed the prompt used to develop the summaries with the initial qualitative review we performed, directing the LLM to focus on entities, calls to action, etc. It is possible that in following the prompt the LLM overemphasized certain elements of a rumor. 

The formulaic nature of the summaries also reduced nuance. Emotion, framing, humor, and sarcasm were often difficult or impossible to recover, making this approach less suitable for research focused on attitudes, reactions, or the specific users involved in discourse. Visual information posed a particular challenge. Rumors driven primarily by images or data visualizations were often obscured in text summaries, potentially leading to undercounting and lower coder consistency for visually communicated tropes. For example, the “F-curve” trope describes visualizations showing a sharp increase in tallied votes followed by a plateau \cite{engelSocialDynamicsMobilization2025}, yet the summaries never used the term “F-curve” and rarely indicated when such visualizations were central to a rumor. Coders therefore sometimes had to rely on word clouds and prior knowledge to infer when the code applied. More generally, this reflects a limitation of language-focused approaches when visual information constitutes a meaningful part of the underlying data. The repetitive structure of the summaries also made it more difficult for coders to sustain attention during long coding sessions.

\begin{acks}
This material is based upon work supported by the National Science Foundation under Grant No.s 2120496 and 1749815. Any opinions, findings, and conclusions or recommendations expressed in this material are those of the authors and do not necessarily reflect the views of the National Science Foundation. This work is also supported by funding from the John S. and James L. Knight Foundation, the William and Flora Hewlett Foundation, and Craig Newmark Philanthropies. The Human Subjects Division determined that the proposed activity does not involve human subjects, as defined by federal and state regulations.
\end{acks}

\bibliographystyle{ACM-Reference-Format}
\bibliography{tropes_references}

\section*{Appendices}
\appendix
\appendix
\section{Full Coding Scheme}
% Requires in preamble (xcolor with [table] must be loaded early):
% \usepackage[table]{xcolor}
% \usepackage{longtable}
% \usepackage{booktabs}
% \usepackage{array}

\definecolor{catBlue}{RGB}{31,119,180}
\definecolor{catGray}{RGB}{110,110,110}
\definecolor{catOrange}{RGB}{230,120,20}
\definecolor{catGreen}{RGB}{44,160,44}
\definecolor{catRed}{RGB}{214,39,40}
\definecolor{catPurple}{RGB}{148,103,189}
\definecolor{catBrown}{RGB}{140,86,75}
\definecolor{catTeal}{RGB}{0,150,170}
\definecolor{catPink}{RGB}{200,80,150}

\setlength{\extrarowheight}{1pt}
\begin{longtable}{>{\raggedright\arraybackslash}p{0.24\textwidth}>{\raggedright\arraybackslash}p{0.28\textwidth}>{\raggedright\arraybackslash}p{0.40\textwidth}}
\caption{Coding scheme of election-fraud tropes.}\label{tab:tropes}\\
\toprule
\rowcolor{gray!25}\textbf{Categories} & \textbf{Meso Level} & \textbf{Micro Level} \\
\midrule
\endfirsthead
\multicolumn{3}{l}{\tablename\ \thetable{} -- continued from previous page}\\
\toprule
\rowcolor{gray!25}\textbf{Categories} & \textbf{Meso Level} & \textbf{Micro Level} \\
\midrule
\endhead
\multicolumn{3}{r}{\emph{Continued on next page}}\\
\endfoot
\bottomrule
\endlastfoot

\cellcolor{catBlue!45}\textbf{Suggested Remedies / Mobilization} & \cellcolor{catBlue!30}\textit{Policy Recommendations} & \cellcolor{catBlue!15}Paper ballots \\
\cellcolor{catBlue!45} & \cellcolor{catBlue!30} & \cellcolor{catBlue!15}Voter ID \\
\cellcolor{catBlue!45} & \cellcolor{catBlue!30} & \cellcolor{catBlue!15}In-person voting - as a policy \\
\cellcolor{catBlue!45} & \cellcolor{catBlue!30} & \cellcolor{catBlue!15}Abolish mail-in voting \\
\cellcolor{catBlue!45} & \cellcolor{catBlue!30} & \cellcolor{catBlue!15}Hand count ballots \\
\cellcolor{catBlue!45} & \cellcolor{catBlue!30} & \cellcolor{catBlue!15}One-day election \\
\cellcolor{catBlue!45} & \cellcolor{catBlue!20}\textit{Call to Action} & \cellcolor{catBlue!8}Audits \\
\cellcolor{catBlue!45} & \cellcolor{catBlue!20} & \cellcolor{catBlue!8}Recounts \\
\cellcolor{catBlue!45} & \cellcolor{catBlue!20} & \cellcolor{catBlue!8}Litigation / lawsuits / prosecution \\
\cellcolor{catBlue!45} & \cellcolor{catBlue!20} & \cellcolor{catBlue!8}Investigations \\
\cellcolor{catBlue!45} & \cellcolor{catBlue!20} & \cellcolor{catBlue!8}Attention (Clicktivism/Slacktivism) - report irregularities \\
\cellcolor{catBlue!45} & \cellcolor{catBlue!20} & \cellcolor{catBlue!8}Military / Police involvement \\
\cellcolor{catBlue!45} & \cellcolor{catBlue!20} & \cellcolor{catBlue!8}Protests \\
\cellcolor{catBlue!45} & \cellcolor{catBlue!20} & \cellcolor{catBlue!8}Armed self-defense at the polls \\
\cellcolor{catBlue!45} & \cellcolor{catBlue!20} & \cellcolor{catBlue!8}``Hold the line'' / ``stay vigilant'' / ``stop the steal'' \\
\cellcolor{catBlue!45} & \cellcolor{catBlue!20} & \cellcolor{catBlue!8}In-person voting - as a call to action \\
\multicolumn{2}{>{\columncolor{catGray!45}\raggedright\arraybackslash}p{0.55\textwidth}}{\textbf{Generic Distrust in Election Procedures}} & \cellcolor{catGray!15}Mail-in voting \\
\multicolumn{2}{>{\columncolor{catGray!45}\raggedright\arraybackslash}p{0.55\textwidth}}{} & \cellcolor{catGray!15}Voting machines \\
\multicolumn{2}{>{\columncolor{catGray!45}\raggedright\arraybackslash}p{0.55\textwidth}}{} & \cellcolor{catGray!15}Voter rolls \\
\cellcolor{catOrange!45}\textbf{Actors} & \cellcolor{catOrange!30}\textit{Heroes} & \cellcolor{catOrange!15}Judges \\
\cellcolor{catOrange!45} & \cellcolor{catOrange!30} & \cellcolor{catOrange!15}Eyewitness/Whistleblower \\
\cellcolor{catOrange!45} & \cellcolor{catOrange!30} & \cellcolor{catOrange!15}Democratic party \\
\cellcolor{catOrange!45} & \cellcolor{catOrange!30} & \cellcolor{catOrange!15}Republican party \\
\cellcolor{catOrange!45} & \cellcolor{catOrange!30} & \cellcolor{catOrange!15}NGO \\
\cellcolor{catOrange!45} & \cellcolor{catOrange!30} & \cellcolor{catOrange!15}Military \\
\cellcolor{catOrange!45} & \cellcolor{catOrange!30} & \cellcolor{catOrange!15}DOJ/FBI/law enforcement \\
\cellcolor{catOrange!45} & \cellcolor{catOrange!30} & \cellcolor{catOrange!15}Patriots \\
\cellcolor{catOrange!45} & \cellcolor{catOrange!30} & \cellcolor{catOrange!15}Other/write in \\
\cellcolor{catOrange!45} & \cellcolor{catOrange!20}\textit{Villains} & \cellcolor{catOrange!8}Big Tech \\
\cellcolor{catOrange!45} & \cellcolor{catOrange!20} & \cellcolor{catOrange!8}Media \\
\cellcolor{catOrange!45} & \cellcolor{catOrange!20} & \cellcolor{catOrange!8}Judges \\
\cellcolor{catOrange!45} & \cellcolor{catOrange!20} & \cellcolor{catOrange!8}Democratic party \\
\cellcolor{catOrange!45} & \cellcolor{catOrange!20} & \cellcolor{catOrange!8}Republican party \\
\cellcolor{catOrange!45} & \cellcolor{catOrange!20} & \cellcolor{catOrange!8}NGO \\
\cellcolor{catOrange!45} & \cellcolor{catOrange!20} & \cellcolor{catOrange!8}Poll workers \\
\cellcolor{catOrange!45} & \cellcolor{catOrange!20} & \cellcolor{catOrange!8}Election officials \\
\cellcolor{catOrange!45} & \cellcolor{catOrange!20} & \cellcolor{catOrange!8}USPS \\
\cellcolor{catOrange!45} & \cellcolor{catOrange!20} & \cellcolor{catOrange!8}Military \\
\cellcolor{catOrange!45} & \cellcolor{catOrange!20} & \cellcolor{catOrange!8}DOJ/FBI/law enforcement \\
\cellcolor{catOrange!45} & \cellcolor{catOrange!20} & \cellcolor{catOrange!8}CIA \\
\cellcolor{catOrange!45} & \cellcolor{catOrange!20} & \cellcolor{catOrange!8}Extremist out-groups \\
\cellcolor{catOrange!45} & \cellcolor{catOrange!20} & \cellcolor{catOrange!8}BLM \\
\cellcolor{catOrange!45} & \cellcolor{catOrange!20} & \cellcolor{catOrange!8}Antifa \\
\cellcolor{catOrange!45} & \cellcolor{catOrange!20} & \cellcolor{catOrange!8}Russia \\
\cellcolor{catOrange!45} & \cellcolor{catOrange!20} & \cellcolor{catOrange!8}China \\
\cellcolor{catOrange!45} & \cellcolor{catOrange!20} & \cellcolor{catOrange!8}Venezuela \\
\cellcolor{catOrange!45} & \cellcolor{catOrange!20} & \cellcolor{catOrange!8}Iran \\
\cellcolor{catOrange!45} & \cellcolor{catOrange!20} & \cellcolor{catOrange!8}Other/write in \\
\cellcolor{catOrange!45} & \cellcolor{catOrange!30}\textit{Victims} & \cellcolor{catOrange!15}Democratic Party / Politicians / Voters \\
\cellcolor{catOrange!45} & \cellcolor{catOrange!30} & \cellcolor{catOrange!15}Republican Party / Politicians / Voters \\
\cellcolor{catOrange!45} & \cellcolor{catOrange!30} & \cellcolor{catOrange!15}Trump campaign / administration \\
\cellcolor{catOrange!45} & \cellcolor{catOrange!30} & \cellcolor{catOrange!15}Biden campaign / administration (includes Harris) \\
\cellcolor{catOrange!45} & \cellcolor{catOrange!30} & \cellcolor{catOrange!15}Minority voters (e.g., African American, Latino) \\
\cellcolor{catOrange!45} & \cellcolor{catOrange!30} & \cellcolor{catOrange!15}Rural voters \\
\cellcolor{catOrange!45} & \cellcolor{catOrange!30} & \cellcolor{catOrange!15}Poll workers \\
\cellcolor{catOrange!45} & \cellcolor{catOrange!30} & \cellcolor{catOrange!15}Election officials \\
\cellcolor{catOrange!45} & \cellcolor{catOrange!30} & \cellcolor{catOrange!15}Law enforcement \\
\cellcolor{catGreen!45}\textbf{Scene and Setting} & \cellcolor{catGreen!30}\textit{Scene (time)} & \cellcolor{catGreen!15}Late night early morning \\
\cellcolor{catGreen!45} & \cellcolor{catGreen!30} & \cellcolor{catGreen!15}COVID-19 Pandemic \\
\cellcolor{catGreen!45} & \cellcolor{catGreen!30} & \cellcolor{catGreen!15}Vote counting (is a time of cheating) \\
\cellcolor{catGreen!45} & \cellcolor{catGreen!30} & \cellcolor{catGreen!15}Election day - near closing polls / evening \\
\cellcolor{catGreen!45} & \cellcolor{catGreen!20}\textit{Setting (place)} & \cellcolor{catGreen!8}Democrat-run city - urban center \\
\cellcolor{catGreen!45} & \cellcolor{catGreen!20} & \cellcolor{catGreen!8}Polling center \\
\cellcolor{catGreen!45} & \cellcolor{catGreen!20} & \cellcolor{catGreen!8}Targeting of Swing States \\
\cellcolor{catGreen!45} & \cellcolor{catGreen!20} & \cellcolor{catGreen!8}Downtown (e.g., offsite counting) \\
\cellcolor{catGreen!45} & \cellcolor{catGreen!20} & \cellcolor{catGreen!8}Absentee/early voting drop box \\
\multicolumn{2}{>{\columncolor{catRed!45}\raggedright\arraybackslash}p{0.55\textwidth}}{\textbf{Mechanism}} & \cellcolor{catRed!15}Elimination (voted but votes discarded) \\
\multicolumn{2}{>{\columncolor{catRed!45}\raggedright\arraybackslash}p{0.55\textwidth}}{} & \cellcolor{catRed!15}Alteration (votes changed) \\
\multicolumn{2}{>{\columncolor{catRed!45}\raggedright\arraybackslash}p{0.55\textwidth}}{} & \cellcolor{catRed!15}Insertion/dilution \\
\multicolumn{2}{>{\columncolor{catRed!45}\raggedright\arraybackslash}p{0.55\textwidth}}{} & \cellcolor{catRed!15}Manipulation (of voters) \\
\multicolumn{2}{>{\columncolor{catRed!45}\raggedright\arraybackslash}p{0.55\textwidth}}{} & \cellcolor{catRed!15}Suppression (stopped from voting) \\
\multicolumn{2}{>{\columncolor{catPurple!45}\raggedright\arraybackslash}p{0.55\textwidth}}{\textbf{Frame Explanations}} & \cellcolor{catPurple!15}Election administration is messy and/or negligent \\
\multicolumn{2}{>{\columncolor{catPurple!45}\raggedright\arraybackslash}p{0.55\textwidth}}{} & \cellcolor{catPurple!15}Untrustworthy results \\
\multicolumn{2}{>{\columncolor{catPurple!45}\raggedright\arraybackslash}p{0.55\textwidth}}{} & \cellcolor{catPurple!15}Elections are vulnerable (to cheating) \\
\multicolumn{2}{>{\columncolor{catPurple!45}\raggedright\arraybackslash}p{0.55\textwidth}}{} & \cellcolor{catPurple!15}Voter fraud or election fraud \\
\multicolumn{2}{>{\columncolor{catPurple!45}\raggedright\arraybackslash}p{0.55\textwidth}}{} & \cellcolor{catPurple!15}Trustworthy elections/frame contesting fraud frames \\
\cellcolor{catBrown!45}\textbf{Actor- Action} & \cellcolor{catBrown!30}\textit{Censorship / Bias} & \cellcolor{catBrown!15}Big Tech \\
\cellcolor{catBrown!45} & \cellcolor{catBrown!30} & \cellcolor{catBrown!15}Media \\
\cellcolor{catBrown!45} & \cellcolor{catBrown!20}\textit{Foreign Interference} & \cellcolor{catBrown!8}Influence operation \\
\cellcolor{catBrown!45} & \cellcolor{catBrown!20} & \cellcolor{catBrown!8}Machine manipulation or cyberattack on election infrastructure \\
\cellcolor{catBrown!45} & \cellcolor{catBrown!20} & \cellcolor{catBrown!8}Manipulated ballots \\
\cellcolor{catBrown!45} & \cellcolor{catBrown!20} & \cellcolor{catBrown!8}Corrupt by association \\
\cellcolor{catBrown!45} & \cellcolor{catBrown!30}\textit{Media error reveals fraud} & \cellcolor{catBrown!15}Media error or glitch (reveals fraud) \\
\cellcolor{catBrown!45} & \cellcolor{catBrown!30} & \cellcolor{catBrown!15}Early media results (reveals fraud) \\
\cellcolor{catBrown!45} & \cellcolor{catBrown!20}\textit{State / district calling the election for a candidate prematurely} & \cellcolor{catBrown!8}Yes/No \\
\cellcolor{catBrown!45} & \cellcolor{catBrown!30}\textit{Slow vote counting} & \cellcolor{catBrown!15}Yes/No \\
\cellcolor{catBrown!45} & \cellcolor{catBrown!20}\textit{Suspicious remedies} & \cellcolor{catBrown!8}Extending voting hours \\
\cellcolor{catBrown!45} & \cellcolor{catBrown!20} & \cellcolor{catBrown!8}Offsite counting \\
\cellcolor{catBrown!45} & \cellcolor{catBrown!20} & \cellcolor{catBrown!8}Placing votes in a Box (Box 3) \\
\cellcolor{catBrown!45} & \cellcolor{catBrown!30}\textit{Voter intimidation} & \cellcolor{catBrown!15}Threats of violence by an (extremist) out-group \\
\cellcolor{catBrown!45} & \cellcolor{catBrown!30} & \cellcolor{catBrown!15}Armed protesters at voting locations \\
\cellcolor{catBrown!45} & \cellcolor{catBrown!30} & \cellcolor{catBrown!15}Armed people at polling station or drop box \\
\cellcolor{catBrown!45} & \cellcolor{catBrown!30} & \cellcolor{catBrown!15}People wearing military gear (camouflage, bullet proof vests) at voting locations \\
\cellcolor{catBrown!45} & \cellcolor{catBrown!30} & \cellcolor{catBrown!15}Police/law enforcement at voting locations \\
\cellcolor{catBrown!45} & \cellcolor{catBrown!30} & \cellcolor{catBrown!15}Domestic intimidation (e.g., family, spouse) \\
\cellcolor{catBrown!45} & \cellcolor{catBrown!30} & \cellcolor{catBrown!15}Anonymous intimidating messages (e.g., calls, texts) \\
\cellcolor{catBrown!45} & \cellcolor{catBrown!20}\textit{USPS sabotage} & \cellcolor{catBrown!8}Slow delivery \\
\cellcolor{catBrown!45} & \cellcolor{catBrown!20} & \cellcolor{catBrown!8}Dumping mail-in ballots \\
\cellcolor{catBrown!45} & \cellcolor{catBrown!30}\textit{Suspicious Non-Action} & \cellcolor{catBrown!15}Frustration with politicians for colluding or not acting (e.g., RINOS) \\
\cellcolor{catBrown!45} & \cellcolor{catBrown!30} & \cellcolor{catBrown!15}Frustration with institutions for colluding or not acting (e.g., DOJ, FBI, AG) \\
\cellcolor{catBrown!45} & \cellcolor{catBrown!20}\textit{Election Official / Poll Worker Negligence / Sabotage} & \cellcolor{catBrown!8}Negligent/incompetent election official (bumbling bureaucrat) \\
\cellcolor{catBrown!45} & \cellcolor{catBrown!20} & \cellcolor{catBrown!8}Malicious election official \\
\cellcolor{catBrown!45} & \cellcolor{catBrown!20} & \cellcolor{catBrown!8}Negligent/incompetent poll worker \\
\cellcolor{catBrown!45} & \cellcolor{catBrown!20} & \cellcolor{catBrown!8}Malicious poll worker \\
\cellcolor{catBrown!45} & \cellcolor{catBrown!30}\textit{Person arrested or charged with real election fraud is evidence of a larger problem} & \cellcolor{catBrown!15}Yes/No \\
\cellcolor{catTeal!45}\textbf{Objects} & \cellcolor{catTeal!30}\textit{Suspicious Containers or Transporters of Ballots} & \cellcolor{catTeal!15}Black duffle bags \\
\cellcolor{catTeal!45} & \cellcolor{catTeal!30} & \cellcolor{catTeal!15}Suitcases \\
\cellcolor{catTeal!45} & \cellcolor{catTeal!30} & \cellcolor{catTeal!15}Containers full of ballots \\
\cellcolor{catTeal!45} & \cellcolor{catTeal!30} & \cellcolor{catTeal!15}White vans \\
\cellcolor{catTeal!45} & \cellcolor{catTeal!30} & \cellcolor{catTeal!15}Unofficial/illegal way of carrying ballots (A) \\
\cellcolor{catTeal!45} & \cellcolor{catTeal!30} & \cellcolor{catTeal!15}Misleading Ballot Dropboxes \\
\cellcolor{catTeal!45} & \cellcolor{catTeal!20}\textit{Suspicious ballot or envelope design} & \cellcolor{catTeal!8}Sharpies/pens (causing problems) \\
\cellcolor{catTeal!45} & \cellcolor{catTeal!20} & \cellcolor{catTeal!8}Holes in envelopes \\
\cellcolor{catTeal!45} & \cellcolor{catTeal!20} & \cellcolor{catTeal!8}Party ID in envelopes \\
\cellcolor{catTeal!45} & \cellcolor{catTeal!20} & \cellcolor{catTeal!8}Candidate ordering on ballot \\
\cellcolor{catTeal!45} & \cellcolor{catTeal!20} & \cellcolor{catTeal!8}Pre-filled ballots \\
\cellcolor{catTeal!45} & \cellcolor{catTeal!30}\textit{Error in ballot preparation (reveals conspiracy)} & \cellcolor{catTeal!15}Typos \\
\cellcolor{catTeal!45} & \cellcolor{catTeal!30} & \cellcolor{catTeal!15}Poorly printed ballots \\
\cellcolor{catTeal!45} & \cellcolor{catTeal!20}\textit{Dead or Ghost Voters} & \cellcolor{catTeal!8}Dead voters \\
\cellcolor{catTeal!45} & \cellcolor{catTeal!20} & \cellcolor{catTeal!8}Ghost voters \\
\cellcolor{catTeal!45} & \cellcolor{catTeal!20} & \cellcolor{catTeal!8}Maidengate Voters \\
\cellcolor{catTeal!45} & \cellcolor{catTeal!20} & \cellcolor{catTeal!8}Pet voters \\
\cellcolor{catTeal!45} & \cellcolor{catTeal!30}\textit{Non-citizen voters} & \cellcolor{catTeal!15}Voting-centered (on the polls) \\
\cellcolor{catTeal!45} & \cellcolor{catTeal!30} & \cellcolor{catTeal!15}Automatic voter registration through driver's licenses \\
\cellcolor{catTeal!45} & \cellcolor{catTeal!20}\textit{Suspicious Statistics / Visualizations} & \cellcolor{catTeal!8}F-Curve \\
\cellcolor{catTeal!45} & \cellcolor{catTeal!20} & \cellcolor{catTeal!8}Improbable Results \\
\cellcolor{catTeal!45} & \cellcolor{catTeal!20} & \cellcolor{catTeal!8}Vote Shifts \\
\cellcolor{catTeal!45} & \cellcolor{catTeal!20} & \cellcolor{catTeal!8}Bullet Ballots \\
\cellcolor{catTeal!45} & \cellcolor{catTeal!20} & \cellcolor{catTeal!8}Split Tickets \\
\cellcolor{catTeal!45} & \cellcolor{catTeal!30}\textit{Voter Suppression through Unfair Rules (complaints about the rules themselves)} & \cellcolor{catTeal!15}Voter suppression through unfair laws, rules, hours and locations of voting \\
\cellcolor{catTeal!45} & \cellcolor{catTeal!30} & \cellcolor{catTeal!15}Long lines \\
\cellcolor{catTeal!45} & \cellcolor{catTeal!30} & \cellcolor{catTeal!15}Gerrymandering \\
\cellcolor{catPink!45}\textbf{Object-Action} & \cellcolor{catPink!30}\textit{Abused Mail-In Ballots} & \cellcolor{catPink!15}Destroyed \\
\cellcolor{catPink!45} & \cellcolor{catPink!30} & \cellcolor{catPink!15}Dumped or discarded or trashed \\
\cellcolor{catPink!45} & \cellcolor{catPink!30} & \cellcolor{catPink!15}Stolen \\
\cellcolor{catPink!45} & \cellcolor{catPink!30} & \cellcolor{catPink!15}Damaged \\
\cellcolor{catPink!45} & \cellcolor{catPink!30} & \cellcolor{catPink!15}Lost or found or discovered \\
\cellcolor{catPink!45} & \cellcolor{catPink!30} & \cellcolor{catPink!15}Missing \\
\cellcolor{catPink!45} & \cellcolor{catPink!20}\textit{Errant Ballot Materials Mailed (reveal loose voter rolls or fraud)} & \cellcolor{catPink!8}Ballot materials mailed to the wrong person \\
\cellcolor{catPink!45} & \cellcolor{catPink!20} & \cellcolor{catPink!8}Multiple ballots or registrations received by the same person \\
\cellcolor{catPink!45} & \cellcolor{catPink!20} & \cellcolor{catPink!8}Ballot materials (including registration) sent to an ineligible person (e.g., non-citizen). \\
\cellcolor{catPink!45} & \cellcolor{catPink!30}\textit{Suspicious (large numbers of) Ballots Found or Inserted} & \cellcolor{catPink!15}Yes/No \\
\cellcolor{catPink!45} & \cellcolor{catPink!20}\textit{Issues with Curing and Vote Auditing} & \cellcolor{catPink!8}Signature matching (e.g., mismatches, verification issues) \\
\cellcolor{catPink!45} & \cellcolor{catPink!20} & \cellcolor{catPink!8}Deleting ballot images \\
\cellcolor{catPink!45} & \cellcolor{catPink!30}\textit{Late ballots counted} & \cellcolor{catPink!15}Yes/No \\
\cellcolor{catPink!45} & \cellcolor{catPink!20}\textit{Oversight Obstruction} & \cellcolor{catPink!8}Cardboard over windows \\
\cellcolor{catPink!45} & \cellcolor{catPink!20} & \cellcolor{catPink!8}Livestreams down \\
\cellcolor{catPink!45} & \cellcolor{catPink!20} & \cellcolor{catPink!8}Blocking poll observers \\
\cellcolor{catPink!45} & \cellcolor{catPink!30}\textit{Disrupted Chain of Custody} & \cellcolor{catPink!15}Yes/No \\
\cellcolor{catPink!45} & \cellcolor{catPink!20}\textit{Ballot Harvesting} & \cellcolor{catPink!8}Ballot trafficking \\
\cellcolor{catPink!45} & \cellcolor{catPink!20} & \cellcolor{catPink!8}Ballot Harvesting \\
\cellcolor{catPink!45} & \cellcolor{catPink!20} & \cellcolor{catPink!8}People paid to vote \\
\cellcolor{catPink!45} & \cellcolor{catPink!20} & \cellcolor{catPink!8}Ballot mules \\
\cellcolor{catPink!45} & \cellcolor{catPink!30}\textit{Malfunctioning Machines} & \cellcolor{catPink!15}Machine ``Glitches'' \\
\cellcolor{catPink!45} & \cellcolor{catPink!30} & \cellcolor{catPink!15}Tabulators/machines ``Down'' \\
\cellcolor{catPink!45} & \cellcolor{catPink!30} & \cellcolor{catPink!15}``Malfunctioning'' Machines \\
\cellcolor{catPink!45} & \cellcolor{catPink!30} & \cellcolor{catPink!15}Machine Vote ``Flips/Switches'' \\
\cellcolor{catPink!45} & \cellcolor{catPink!20}\textit{Manipulated Machines} & \cellcolor{catPink!8}Hacked Machines (manipulated by an outsider) \\
\cellcolor{catPink!45} & \cellcolor{catPink!20} & \cellcolor{catPink!8}Corrupt Machines (designed to manipulate) \\
\cellcolor{catPink!45} & \cellcolor{catPink!30}\textit{Suspicious Disruption to In-Person Voting} & \cellcolor{catPink!15}Closures \\
\cellcolor{catPink!45} & \cellcolor{catPink!30} & \cellcolor{catPink!15}Bomb threats \\
\cellcolor{catPink!45} & \cellcolor{catPink!30} & \cellcolor{catPink!15}Voting facility issues (Power down, floods, disasters) \\
\end{longtable}

\section{Shortened Label Mapping}
% Required packages: \usepackage{booktabs} \usepackage{longtable}

\begin{longtable}{p{0.55\textwidth} p{0.35\textwidth}}
\caption{Code Label Abbreviations Used in Network Visualizations}
\label{tab:code_labels} \\
\toprule
\textbf{Original Label} & \textbf{Shortened Label} \\
\midrule
\endfirsthead
\multicolumn{2}{c}{\tablename\ \thetable{} -- \textit{continued from previous page}} \\
\toprule
\textbf{Original Label} & \textbf{Shortened Label} \\
\midrule
\endhead
\midrule
\multicolumn{2}{r}{\textit{Continued on next page}} \\
\endfoot
\bottomrule
\endlastfoot
    Abolish mail-in voting & No mail-in voting \\
    Absentee/early voting drop box & Voting drop box \\
    Anonymous intimidating messages (e.g., calls, texts) & Anonymous threats \\
    Armed self-defense at the polls & Armed at polls \\
    Attention (Clicktivism/Slacktivism) - report irregularities & Online activism \\
    Automatic voter registration through driver's licenses & Auto voter registration \\
    Ballot Harvesting & Ballot harvesting \\
    Ballot materials mailed to the wrong person & Misdelivered ballots \\
    Ballot mules & Ballot mules \\
    Ballot trafficking & Ballot trafficking \\
    Biden campaign / administration (includes Harris) & Biden admin \\
    Black duffle bags & Duffle bags \\
    Bullet Ballots & Bullet ballots \\
    COVID-19 Pandemic & COVID pandemic \\
    Cardboard over windows & Covered windows \\
    Chain of custody disruption / loss & Chain of custody loss \\
    Containers full of ballots & Containers of ballots \\
    Corrupt Machines & Corrupt machines \\
    Corrupt by association & Corrupt by association \\
    Deleting ballot images & Deleted ballot images \\
    Democrat-run city - urban center & Dem-run city \\
    Democrat-run city - urban center; Polling center & Polling center \\
    Polling center & Polling center \\
    Democratic Party / Politicians / Voters & Democrats \\
    democratic party & Democrats \\
    Domestic intimidation (e.g., family, spouse) & Family intimidation \\
    Downtown (e.g., offsite counting) & Downtown \\
    Dumping mail-in ballots & Dumped mail-in ballots \\
    Early media results (reveals fraud) & Early media results \\
    Election day - near closing polls / evening & Election night \\
    Election officials & Election officials \\
    Extending voting hours & Extended voting hours \\
    Extremist out-groups & Extremist groups \\
    Eyewitness/Whistleblower / contrarian expert & Whistleblower \\
    Frustration with institutions for colluding or not acting (e.g., DOJ, FBI, AG) & Institutional non-action \\
    Frustration with politicians for colluding or not acting (e.g., RINOS) & Politician collusion \\
    Gerrymandering & Gerrymandering \\
    Hacked Machines & Hacked machines \\
    Hand count ballots & Hand count \\
    Improbable Results & Improbable results \\
    In-person Voting - as a call to action & Vote in person (action) \\
    In-person Voting - as a policy & Vote in person (policy) \\
    Influence operation & Influence operation \\
    Investigations & Investigations \\
    Late night early morning & Late night/early morning \\
    Litigation / lawsuits / prosecution & Litigation \\
    Livestreams down & Livestreams down \\
    Machine Glitches & Machine glitches \\
    Machine Vote Flips/Switches & Machine vote flip \\
    Machine manipulation or cyberattack on election infrastructure & Machine manipulation \\
    Maidengate Voters & Maidengate voters \\
    Malicious election official & Malicious official \\
    Malicious poll worker & Malicious poll worker \\
    Manipulated ballots & Manipulated ballots \\
    Media error or glitch (reveals fraud), & Media error \\
    Media figure - influencer & Media influencer \\
    Media figure / influencer & Media influencer \\
    Military / Police involvement & Military/police \\
    Minority voters (e.g., African American, Latino) & Minority voters \\
    Multiple ballots or registrations received by the same person & Multiple ballots received \\
    Negligent/incompetent election official & Negligent official \\
    Negligent/incompetent election official (bumbling bureaucrat) & Negligent official \\
    Negligent/incompetent poll worker & Negligent poll worker \\
    Offsite counting & Offsite counting \\
    Paper ballots & Paper ballots \\
    People paid to vote & Paid voters \\
    Placing votes in a Box (Box 3) & Backup ballot box \\
    Pre-filled ballots & Pre-filled ballots \\
    Republican Party / Politicians / Voters & Republicans \\
    republican party / politicians / voters & Republicans \\
    Sharpies/pens (causing problems) & Sharpie issues \\
    sharpies/pens (causing problems) & Sharpie issues \\
    Signature matching (e.g., mismatches, verification issues) & Signature mismatch \\
    Signatures not matching & Signature mismatch \\
    Slow delivery & Slow delivery \\
    Split Tickets & Split tickets \\
    Targeting of Swing States & Swing states \\
    Targeting of swing states & Swing states \\
    Threats of violence by an (extremist) out-group & Extremist threats \\
    Trump / Campaign / Administration & Trump admin \\
    Trump campaign / administration & Trump admin \\
    Unofficial/illegal way of carrying ballots (A) & Unofficial ballot transport \\
    Violent Protests / Riots & Violent protests \\
    Vote counting (is a time of cheating) & Vote counting \\
    Voter suppression through unfair laws, rules, hours and locations of voting & Voter suppression (rules) \\
    Voting-centered & Noncitizen voters \\
    armed people at polling station or drop box & Armed at polls \\
    blocking poll observers & Blocking observers \\
    dumped or discarded or trashed & Dumped ballots \\
    hold the line & Hold the line \\
    hold the line - vigilance -stop the steal & Hold the line/stop steal \\
    holes in envelopes & Holes in envelopes \\
    internet outages & Internet outages \\
    lost or found or discovered & Found ballots \\
    party ID in envelopes & Party ID visible \\
    people wearing military gear (camouflage, bullet proof vests) & Military gear at polls \\
    poorly printed ballots & Poorly printed ballots \\
    voting facility issues (Power down, floods, disasters) & Facility issues \\
    DOJ/FBI/law enforcement & Law enforcement \\
    Law enforcement & Law enforcement \\
    Police/law enforcement & Law enforcement \\
    Democratic party & Democrats \\
    Republican party & Republicans \\
    Malfunctioning Machines & Malfunctioning machines \\
    Tabulators Down & Tabulators down \\
    Suitcases & Suitcases \\
    white vans & White vans \\
    Antifa & Antifa \\
    BLM & BLM \\
    China & China \\
    Deep State & Deep State \\
    Elderly voters & Elderly voters \\
    Iran & Iran \\
    Judges & Judges \\
    Media & Media \\
    Military & Military \\
    NGO & NGO \\
    Poll workers & Poll workers \\
    Proud Boys & Proud Boys \\
    Russia & Russia \\
    USPS & USPS \\
    Venezuela & Venezuela \\
    CIA & CIA \\
    Recounts & Recounts \\
    Audits & Audits \\
    Protests & Protests \\
    Patriots & Patriots \\
    Dead voters & Dead voters \\
    Ghost voters & Ghost voters \\
    Pet voters & Pet voters \\
    Typos & Typos \\
    damaged & Damaged ballots \\
    destroyed & Destroyed ballots \\
    missing & Missing ballots \\
    stolen & Stolen ballots \\
    F-Curve & F-Curve \\
    Vote Shifts & Vote shifts \\
    Closures & Closures \\
    Long lines & Long lines \\
    Voter ID & Voter ID \\
    Bomb threats & Bomb threats \\
    bomb threats & Bomb threats \\
    Election officials & Election officials \\
\end{longtable}

\end{document}